\documentclass[11pt]{article}
\newlength{\defbaselineskip}
\usepackage{amssymb}
\usepackage{amsfonts}
\usepackage{amsmath}
\usepackage{graphicx}
\usepackage{url}
\usepackage{rotating}
\usepackage{multirow}
\usepackage{color}
\usepackage{xcolor}
\usepackage{enumitem}

\usepackage{subfigure}
\usepackage{xcolor}
\usepackage{adjustbox}
\usepackage{algorithm, algorithmic}
\usepackage{dsfont}
\usepackage[scientific-notation=true]{siunitx}
\definecolor{darkgreen}{rgb}{0,0.6,0}

\usepackage[multiple]{footmisc}

\begin{document}

\title{LASAGNE: Locality And Structure Aware Graph Node Embedding}

\author{Evgeniy Faerman\thanks{Ludwig-Maximilians-Universit\"at M\"unchen}
\and
Felix Borutta\footnotemark[1]
\and
Kimon Fountoulakis\thanks{University of California at Berkeley}
\and
Michael W. Mahoney\footnotemark[2]
}

\date{}
\maketitle

\begin{abstract}
\noindent
Recent work has attempted to identify structure in social and information graphs by using the following approach: first, use random walk methods to explore the neighborhood of a node; second, use ideas from natural language processing to use this neighborhood information to learn vector representations of these nodes reflecting properties of the graph.
Informally, the idea is that if a node is a member of a meaningful cluster or community, then the vector representation should be higher-quality, thereby leading to improved learning.

In this paper, we identify and remedy an important shortcoming of this approach.
In particular, we show that the performance of existing methodologies depends strongly on the structural properties of the graph, e.g., the size of the graph, whether the graph has a flat or upward-sloping Network Community Profile (NCP), whether the graph is expander-like, whether the classes of interest are more $k$-core-like or more peripheral, etc. 
For larger graphs with flat NCPs that are strongly expander-like, existing methods lead to random walks that expand rapidly, touching many dissimilar nodes, thereby leading to lower-quality vector representations that are less useful for downstream tasks.

Based on our findings, we propose \textsc{Lasagne}, a methodology to learn locality and structure aware graph node embeddings in an unsupervised way. 
Rather than relying on global random walks or neighbors within fixed hop distances, \textsc{Lasagne} exploits strongly local \textit{Approximate Personalized PageRank} stationary distributions to more precisely engineer local information into node embeddings.
This leads, in particular, to more meaningful and more useful vector representations of nodes in poorly-structured graphs.
We show that \textsc{Lasagne} leads to significant improvement in downstream multi-label classification for larger graphs with flat NCPs, that it is comparable for smaller graphs with upward-sloping NCPs, and that is comparable to existing methods for link prediction tasks.
\end{abstract}

\section{Introduction}
Graphs are a common way to describe interactions between entities. 
The entities are modeled as nodes, and the interactions between pairs of entities are represented by edges between nodes. 
Describing nodes of a graph as low dimensional vectors has the advantage that many popular machine learning algorithms can be automatically applied, and it is applicable in many areas like visualization, link prediction, classification, etc.~\cite{maaten2008visualizing, liben2007link, aggarwal2011introduction, bhagat2011node}. 
Motivated by this, so-called representation learning methods for graph vertices, e.g.,~\cite{perozzi2014deepwalk, tang2015line, grover2016node2vec}, focus on learning vectors to represent information in neighborhoods around a node, e.g., nodes within a short geodesic distance or nodes encountered in random walks starting at a given node. 

Somewhat more formally, let $G=(V,E)$ be a graph, with $V=\lbrace v_1, \ldots v_N \rbrace$ being the set of nodes and $E=\lbrace e~\vert e \in V \times V \rbrace$ being the set of (undirected) edges. 
The general goal is to find a vector embedding or latent representation for each node $v_i$ such that the resulting set of embedded nodes $\mathcal{E}=\lbrace f(v_i) \vert v_i \in V \rbrace$ in the $d$-dimensional vector space $\mathds{R}^d$ still reflects structural properties of $G$. 
For instance, such structural properties could be the similarity of the neighborhoods of two nodes $v_i$ and $v_j$. 
The neighborhood $\mathcal{N}(v)$ of a node $v$ is defined as the set of nodes having the highest probabilities to be visited by a random walk starting from node $v$, a geodesic walk starting from $v$, or some other related process. 
This means if $\mathcal{N}(v_i) \approx \mathcal{N}(v_j)$ holds in the original graph, it should also hold that $f(v_i) \approx f(v_j)$ in $\mathds{R}^d$.

The intuition behind these representation learning methods is that nodes having similar neighborhoods are similar to each other, and thus one can use information in the neighbors of a node to make predictions for a given node. 
Defining the right neighborhood for each node, however, is a challenging task.
For example, in unsupervised multi-label classification, the labels of the nodes define the underlying local structure for a particular class, but often this does not necessarily overlap significantly with the local structure defined by the edge connectivity of the graph. 
Alternatively, realistic graphs typically have large-scale properties that are very poorly structured with respect to the behavior of random walks \cite{LLDM08_communities_CONF,leskovec2009community,LLM10_communities_CONF,jeub2015think,ASM13,ASM16_IM}.

The basic assumption of random walk based methods and, of course, the large body of very related methods based on spectral graph theory is that nodes visited more often than others by random walks starting from a particular node are also more useful to describe that node in terms of downstream prediction tasks. 
However, the problem with random walks is that typically most of the graph can be reached within a few steps, and thus information about where the random walk began (which is the node for which these methods are computing the embedding) is quickly lost.

This issue is particularly problematic for extremely sparse graphs with upward-sloping Network Community Profiles (NCPs)~\cite{LLDM08_communities_CONF,leskovec2009community,LLM10_communities_CONF} and for flat NCPs~\cite{jeub2015think} (expander-like graphs) or deep $k$-cores~\cite{ASM13,ASM16_IM}.
These properties are ubiquitous among realistic social and information networks.
This suggests that, unless carefully engineered, embedding methods based on random walks will perform sub-optimally, since the random walks will mix rapidly, thereby degrading the local information that one hopes they identify.

In this work, we explore these issues, and we present a method which takes into account the local neighborhood structure of each node in the graph individually. 
This leads to insight into how to better exploit graph topology in poorly structured graphs, and it can result in improved embedding vectors.

Our method, \textsc{Lasagne}, is an unsupervised algorithm for learning locality and structure aware graph node embeddings.
It uses an \textit{Approximate Personalized PageRank} vector~\cite{andersen2006local} to adapt and improve state-of-the-art methods for determining the importance of the nodes in a graph from a specific node's point of view. 
The proposed methodology is easily parallelizable, even on distributed environments, and it has even been shown that the methods we adapt were applied to graphs with more than billions of nodes on a single machine \cite{Shun:2016:PLG:2994509.2994522}.

We evaluate our algorithm with multi-label classification and link prediction on several real-world datasets from different domains under real-life conditions. 
Our evaluations show that our algorithm achieves better results---especially for downstream machine learning tasks whose objectives are sensitive to local information---in terms of prediction accuracy than the state-of-the-art methods, and our algorithm achieves similar results for link prediction. 
As has been described previously~\cite{leskovec2009community,jeub2015think,ASM13,ASM16_IM}, and as we review in Section~\ref{sec:experiments}, graphs with flat NCPs and many deep $k$-core nodes have local structure that is particularly difficult to identify and exploit.
Importantly, our empirical results for this class of graphs is substantially improved, relative to previous methods.
This illustrates that, by carefully engineering locally-biased information into node embeddings, one can obtain improved results even for this class of graphs, without sacrificing quality on other less poorly-structured graphs.

We also illustrate several reasons why random walk based methods do not perform as expected in practice, justifying our interpretation that our method leads to improved results due to the manner in which we engineer in locality. 

The remainder of the paper is as follows: 
in Section \ref{sec:preliminaries}, we survey related work, including the \textit{word2vec} framework and the approximate computation of the \textit{Personalized PageRank}; 
in Section~\ref{sec:mainpart}, we describe our main \textsc{Lasagne} algorithm; 
in Section~\ref{sec:experiments}, we present the evaluation of our method and a discussion of disadvantages of previous random walk based methods; and 
in Section~\ref{sec:conclusion}, we present a brief conclusion.

\section{Preliminaries} 
\label{sec:preliminaries}

\subsection{Related Work on Node Embedding}
\label{sec:relatedwork}
There has been a large body of work on using global spectral methods to compute embeddings for the nodes of a graph for use in machine learning and data analysis problems~\cite{tenenbaum2000global,roweis2000nonlinear,belkin2001laplacian}.
More related is recent representation learning methods for graph vertices that try to construct vectors on the basis of local neighborhood information~\cite{perozzi2014deepwalk, tang2015line, cao2015grarep, grover2016node2vec}. 

The unsupervised \textit{DeepWalk} algorithm learns latent representations for graph vertices by using multiple random walks~\cite{perozzi2014deepwalk}; and it then applies the \textit{Skip-gram} model, originating from natural language processing, to the sequences of nodes given by the random walks. 
The \textit{LINE} algorithm learns two different representations~\cite{tang2015line}, the first of which encourages two nodes to have close embeddings when they are directly connected, and the second of which encourages two nodes to be close when they share the same direct neighbors. 
The \textit{GraRep} algorithm takes this a step further and computes a sequence of matrices, random walk transition matrices taken to powers ranging from $1$ to $k$, and it then applies the SVD to them~\cite{cao2015grarep}. 
Most recently, Grover et al. presented the so-called \textit{node2vec} method~\cite{grover2016node2vec}, a semi-supervised method that borrows the idea of \textit{word2vec} (from natural language processing) to learn node embeddings. 
Instead of using a random search strategy, \textit{node2vec} introduces two hyperparameters to use second order random walks in order to bias the random walks towards a particular search strategy.

Finally, we should note that Yang et al. proposed a semi-supervised learning technique which 
combines information from local neighborhood with information about class labels \cite{yang2016revisiting}. 
The learning algorithm alternates between prediction of nodes in neighborhood as in \textit{DeepWalk} and prediction of nodes having the same labels.
Further related approaches can be found in \cite{scarselli2009graph,yan2005graph,li2014lrbm, li2016gated, tian2014learning,kipf2016semi}.
\subsection{Embedding Words with Word2vec}
\label{sec:word2vec}

\textit{Word2vec} \cite{mikolov2013efficient,le2014distributed} is a framework for learning word representations in some vector space by simultaneously preserving the words' semantic meaning. The representations are learned based on some contexts so that  embeddings sharing similar contexts end up close to each other in the learned space.  
The embeddings are learned  by maximizing the prediction probability of the contexts given the input embeddings, i.e., \textit{Skip-gram} model.
Note, that the model assumes independence of different contexts from each other for the same input.
\textit{Negative sampling} is used to estimate the prediction probability during the training.
It maximizes the log probability of the input's context by simultaneously minimizing the prediction probability for $k$ randomly selected contexts.
Furthermore logistic regression is used to estimate the prediction probability:
\vspace{-2ex}  
$$ \log \sigma(v_{I}'^T v_{c_{i}}) + \sum_{j=1}^k \mathds{E}_{w_{j\sim P_{n}(w)}} \log \sigma(-v_{I}'^T v_{j}).
\vspace{-1ex}$$
For each word the model maintains two representations, embedding and context representation.
The vector $v'_{I}$ denotes the embedding representation of the input, $v_{c_{i}}$ is the context representation and $v_{j}$ are representations of randomly selected contexts.
The stochastic gradient descent algorithm is used for model optimization.
An analysis of this \textit{word2vec} method has been provided by~\cite{gittens2017skip}, reflecting a perspective similar to ours.
The \textit{Skip-gram} was later generalized to include arbitrary contexts \cite{levy2014dependency}.
\subsection{Approximate Personalized PageRank}
The \textit{PageRank} algorithm \cite{page1999pagerank} computes an ``importance'' score for every node in some graph.
Each of the scores corresponds to the probability of a ``random surfer" to visit a node given some start distribution.
The \textit{PageRank} vector is the solution of the linear system:
\begin{equation} \label{eq:basic_pagerank}
pr(s) = \alpha s + (1-\alpha)pr(s)W,
\vspace{-1ex}
\end{equation}
with $W = D^{-1}A$ being the random walk transition matrix. 
$A$ is the adjacency matrix, $D$ is the degree matrix having the node degrees on the diagonal.
The constant $\alpha$ is the \textit{teleportation} probability.
The starting nodes or more specifically the probability for each node to be the starting point of a random walk are given by the vector $s$.
A variant of \textit{PageRank} is the \textit{Personalized PageRank} (PPR) whose result corresponds to the result of the \textit{PageRank} algorithm, where the probabilities in the starting vector $s$ are biased towards some set of nodes. 
The \textit{push} algorithm described in \cite{jeh2003scaling} \cite{berkhin2006bookmark} \cite{andersen2006local} is used  to compute an \textit{Approximate Personalized PageRank} (APPR) vector in a more efficient way if the start distribution vector $s$ is sparse, i.e., has probability mass on only a few nodes.
The idea behind the \textit{push} algorithm is to propagate a node's probability locally and only if there is a sufficient amount of probability to update.
This leads to a sparse solution which means that only relatively few nodes of the underlying graph are contained in the resulting APPR vector.%
\footnote{We emphasize that this APPR method has been remarkably successful at characterizing the local and global structural properties in large social and information networks
\cite{LLDM08_communities_CONF,leskovec2009community,LLM10_communities_CONF,jeub2015think}, suggesting (as we show here) that it can also be used for improved supervised learning on these graphs.}

We describe the adapted version from \cite{Shun:2016:PLG:2994509.2994522} which converges faster.
In addition to $\alpha$ and $s$, the main algorithm expects the approximation parameter $\epsilon$.
It maintains two vectors: the solution vector $p$ and a residual vector $r$.
The $p$ vector is the current approximation of the PPR vector and vector $r$ contains the approximation error or not yet distributed probability mass.
In each iteration the main algorithm selects a node with sufficient probability mass in vector $r$ and calls the \textit{push} method.
The probability mass from node's entry in $r$ is spread  between the node entry in $p$ and the entries of its direct neighbors in $r$.
The exact PPR is the linear combination of the current solution vector $p$ and the PPR solution for $r$. %
The approximation quality and runtime are controlled by the parameter $\epsilon$.
The updates are performed as long as there is a node for which at least $\epsilon \frac{1-\alpha}{1+\alpha}$ probability mass is moved towards each of its neighbors during the \textit{push} operation.

\section{LASAGNE: Locality And Structure Aware Graph Node Embedding}
\label{sec:mainpart}
The \textsc{Lasagne} algorithm consists of two steps: a preprocessing step, which computes the APPR vectors for each node; and the learning step, which uses the APPR vectors to generate training examples batchwise to learn the final embeddings.%

\subsection{Approximated Personalized PageRank for Node Embeddings}
The computation of the APPR vectors for the node embeddings is described in Algorithm \ref{alg:apprForNodeEmbeddings}. 
\begin{algorithm}[t]
	\algsetup{linenosize=\fontsize{10}{11.6}}
	\fontsize{10}{11.6}
	\selectfont
	\caption{\textsc{Lasagne} ApproximatePPR}\label{alg:apprForNodeEmbeddings}
	\begin{algorithmic}[1]
		\renewcommand{\algorithmicrequire}{\textbf{Input:}}
		\renewcommand{\algorithmicensure}{\textbf{Output:}}
		\REQUIRE Node $s$, teleportation parameter $\alpha$, Probability significance threshold $\delta$
		\ENSURE APPR vector p
		\STATE p = $\vec{0}$, r = $\vec{0}$, heap=heap()
		\STATE r(s) = 1
		\STATE heap.push((s,1))
		\STATE sumProbUpdates = 0
		\STATE lastDistrUpdate = 1
		\WHILE {lastDistrUpdate $>$ $\delta$}
			\STATE u = heap.pop()
			\STATE probUpdate = $(2\alpha/(1+\alpha))$r(u)
			\IF {u $\neq$ s}
				\STATE sumProbUpdates += probUpdate
				\STATE lastDistrUpdate = probUpdate / sumProbUpdates
			\ENDIF
			\STATE p(u) = p(u) + probUpdate
			\STATE neighResUpdate = $((1-\alpha)/(1+\alpha))r(u)/d(u)$
			\FOR {v with (u,v)$\in E$}
			\STATE r(v) = r(v) + neighResUpdate
			\STATE heap.update((v, r(v)/size(v.neighbours)))
		\ENDFOR
		\STATE r(u) = 0
		\ENDWHILE
		\STATE p(s) = 0 \label{alg:apprForNodeEmbeddings:replOwnProbStart}
		\STATE p(s) = max(p)\label{alg:apprForNodeEmbeddings:replOwnProbEnd}
		\RETURN p
		\medskip
	\end{algorithmic}
\end{algorithm}
There are two main modifications, relative to the original method in \cite{Shun:2016:PLG:2994509.2994522}.
The first is the assignment of probability mass to the \textit{seed node} in its own APPR vector, and the second is to the stopping criterion.%
\footnote{These modifications seem minor, but getting them right is extremely important for obtaining a robust and successful method.}

The first modification allows the seed node to be considered as its own neighbor during sampling the training examples. 
Consequently, the seed node is considered to be similar to other nodes that have the seed node among their neighbors, which in turn leads to higher proximity of such nodes in the embedded space. 
To avoid each node being considered to be the most important member of its own neighborhood (and thus being overrepresented during the training phase), we replace the node's own entry in its APPR vector with the second highest probability, c.f., line \ref{alg:apprForNodeEmbeddings:replOwnProbStart} - \ref{alg:apprForNodeEmbeddings:replOwnProbEnd}. 

The second modification is since 
our main motivation is not to approximate the PPR vector but instead to keep only the \textit{relevant} neighbors that represent a meaningful context for the seed node. 
The algorithm avoids considering neighbors, each of which is visited relatively rarely by the random walk.%
\footnote{These nodes tend to be ``far from'' the node of interest; but, in total, they may absorb a significant large amount of the overall probability mass.}
Thus, our algorithm stops when the new node, which can be added to the APPR vector during the next iteration has a low chance to be visited by the random walk compared to the overall probability of previously added nodes. 
The running time for the algorithm depends on the probability significance threshold $\delta$. The number of updates of the APPR vector is at most $\frac{1}{\delta}$. Given that the amount of probability moved in subsequent steps is always lower, we can assume it to be the same. Therefore it holds that $sumProbUpdates = n\cdot probUpdate$, whit $n$ being the number of previous steps. Given that $lastDistrUpdate = probUpdate /sumProbUpdates$, it follows that $lastDistrUpdate \leq \frac{1}{n}$.

\subsection{Learning of Embeddings From Approximated Personalized PageRank Vector}
The embedding learning process is described in Algorithm \ref{alg:learnEmbeddings}.
\begin{algorithm}[t]
	\algsetup{linenosize=\fontsize{10}{11.6}}
	\fontsize{10}{11.6}
	\selectfont
	\caption{Learn Embeddings}\label{alg:learnEmbeddings}
	\begin{algorithmic}[1]
		\renewcommand{\algorithmicrequire}{\textbf{Input:}}
		\renewcommand{\algorithmicensure}{\textbf{Output:}}
		\REQUIRE List with seed node and APPR vector pairs \textit{apprs}, \textit{maxBatches}, \textit{batchSize}
		\STATE samplers = emptyList()
		\FOR {seedNode and currentAppr \textbf{in} \textit{apprs}}
			\STATE samplers.add((seedNode, createAliasSampler(currentAppr))
		\ENDFOR
		\WHILE {not converged and batchNumber $<$ \textit{maxBatches}}
			\STATE currentBatch = emptyList()
				\FOR {seedNode and s \textbf{in} samplers}
				\STATE neighbors = s.sample(\textit{batchSize} / size(samplers))
				\STATE trainingExamples = createPairs(seedNode, neighbors)
				\STATE currentBatch.add(trainingExamples)
				\ENDFOR
			\STATE permute (currentBatch)
			\STATE negativeSamplingGradientDescent(currentBatch)
			\STATE batchNumber++

		\ENDWHILE
		\medskip
	\end{algorithmic}
\end{algorithm}
Each training example is a pair of nodes.  We call one of them \textit{seed node} and the other one \textit{neighbor node}. The embedding is learned for the \textit{seed node} while the \textit{neighbor node} is used as context. 
The embeddings are learned analogously to the \textit{Skip-gram} model described in Section \ref{sec:word2vec}. For each training pair the probability of the \textit{neighbor node} is maximized given the \textit{seed node}. 

To generate the training pairs, we sample the \textit{neighbor nodes} based on the APPR vector of the corresponding \textit{seed node}. This means that for each \textit{seed node}, we consider only those nodes as context which have some probability mass in the \textit{seed node's} APPR vector, i.e., relevant nodes.
Each \textit{neighbor node}  is sampled with the probability proportionally to its entry in the \textit{seed's} APPR vector.
\textit{Neighbor nodes} are sampled with replacement and the probability to be sampled is equal to the relative ratio of probability mass each \textit{neighbor node} contributes to the entire APPR vector.
With this sampling strategy training data can be generated on request and the number of training examples per node can be easily controlled, and it leads to higher quality training data.
Using the alias method \cite{knuth1969art}, the sampling setup costs are $O(k)$, where $k$ is the size of the APPR vector and the costs to sample a \textit{neighbor} are $O(1)$.

\subsection{Parallelization}
Our approach scales linearly with number of nodes and can easily be parallelized. 
The APPR vectors can be computed independently for each node and as shown in \cite{Shun:2016:PLG:2994509.2994522} even the largest publicly available graphs fit into the memory of todays commodity hardware. 
The learning procedure can be parallelized in two ways: the sampling from APPR can be done independently in parallel; and the actual learning of the embeddings can also be processed in parallel either asynchronously or synchronously on multi-core or distributed architectures. 
For details see \cite{ji2016parallelizing}.

\section{Empirical results}
\label{sec:experiments}
In this section, we summarize our empirical results.
We have evaluated the node embeddings produced by the \textsc{Lasagne} algorithm by performing prediction tasks which aim at inferring node labels in multi-label classification and link prediction scenarios. 
We have used a variety of real-world graph datasets from various domains, i.e., a biological network, social networks, and a collaboration network. 
Here, we compare our results against the state-of-the-art techniques \textit{DeepWalk}, \textit{node2vec} and \textit{GraRep}. Note that we omit a comparison with the \textit{LINE} since it is already shown in \cite{grover2016node2vec} and \cite{cao2015grarep} that the results produced by \textit{node2vec} and \textit{GraRep} are superior to the ones produced by \textit{LINE}. We have implemented \textit{GraRep} using sparse matrix operations. 
Despite of this, we were not able to run it for larger graphs due to out of memory errors. 
We tested on a machine with 387GB RAM. 

\subsection{Datasets}
\begin{table}[htb]
\begin{center}
\begin{adjustbox}{max width=\textwidth}
  \begin{tabular}{ | l | c | c | c | c | c | c | c | c | c | c |}
    \hline
    Network & $\vert V\vert$ & $\vert E\vert$ & $\vert \mathcal{L} \vert$ & $\overline{d}$ & $\overline{C}$ & $D$ & $\overline{D}$ & $ k_{max}$ & $P_{k_{max}}$ & Description \\ \hline
    PPI & 3,890 & 38,739 & 50 & 9.959 & 0.146 & 8 & 3.1 & 30 & 0.028 & biological network \\ \hline
    BlogCatalog & 10,312 & 333,983 & 39 & 32.388 & 0.463 & 5 & 2.4 & 115 & 0.043 & social network \\ \hline
    IMDb Germany & 32,732 & 1,175,364 & 27 & 35.909 & 0.870 & 11 & 3.5 & 102 & 0.009 & collaboration network \\ \hline
    Flickr & 80,513 & 5,899,882 & 195 & 73.279 & 0.165 & 6 & 2.9 & 551 & 0.018 & social network \\ \hline   
  \end{tabular}
  \end{adjustbox}
  \end{center}
  \caption{Statistics of networks used for multi-label classification: number of nodes $\vert V\vert$, number of edges $\vert E\vert$, number of classes $\vert \mathcal{L} \vert$, average degree $\overline{d}$, average clustering coefficient $\overline{C}$, diameter $D$ and average shortest path length $\overline{D}$, maximum $k$ of $k-cores$ $k_{max}$, fraction $P_{k_{max}}$ of nodes in $k_{max}$ $k-core$}\label{tab:statistics}
\end{table}
We consider the following graph datasets from various domains with different sizes and number of classes.
\begin{itemize}
\item Protein-Protein Interactions (PPI) \cite{breitkreutz2008biogrid}: This is a subgraph of the PPI network for Homo Sapien{}s which is also used in \cite{grover2016node2vec}. The network consists of 3,890 nodes that represent proteins and 38,739 edges which represent the existence of interactions between the corresponding proteins. The 50 different labels represent biological states.
\item BlogCatalog \cite{tang2009relational}: This is a social network graph where each of the 10,312 nodes corresponds to a user and the 333,983 edges represent the friendship relationships between bloggers. 39 different interest groups provide the labels. This network is used in both \cite{grover2016node2vec} and \cite{perozzi2014deepwalk}.
\item IMDb Germany: This kind of artificial dataset is created from the IMDb movie database \cite{imdb2016the}. It consists of 32,732 nodes, 1,175,364 edges and 27 labels. Each node represents an actor/actress who played in a German movie. Edges connect actors/actresses that were in a cast together and the node labels represent the genres that the corresponding actor/actress played. 
\item Flickr \cite{tang2009relational}: The Flickr network is a social network graph with 80,513 nodes and 5,899,882 edges. Each node describes a user and the links represent friendships. The 195 given labels stem from different interest groups. This dataset is also used in \cite{perozzi2014deepwalk}.
\end{itemize}
Table~\ref{tab:statistics}  summarizes some statistics of these networks.

\begin{figure}[t]
\begin{center}
    \subfigure[PPI]{
        \includegraphics[width =
        0.45\columnwidth]{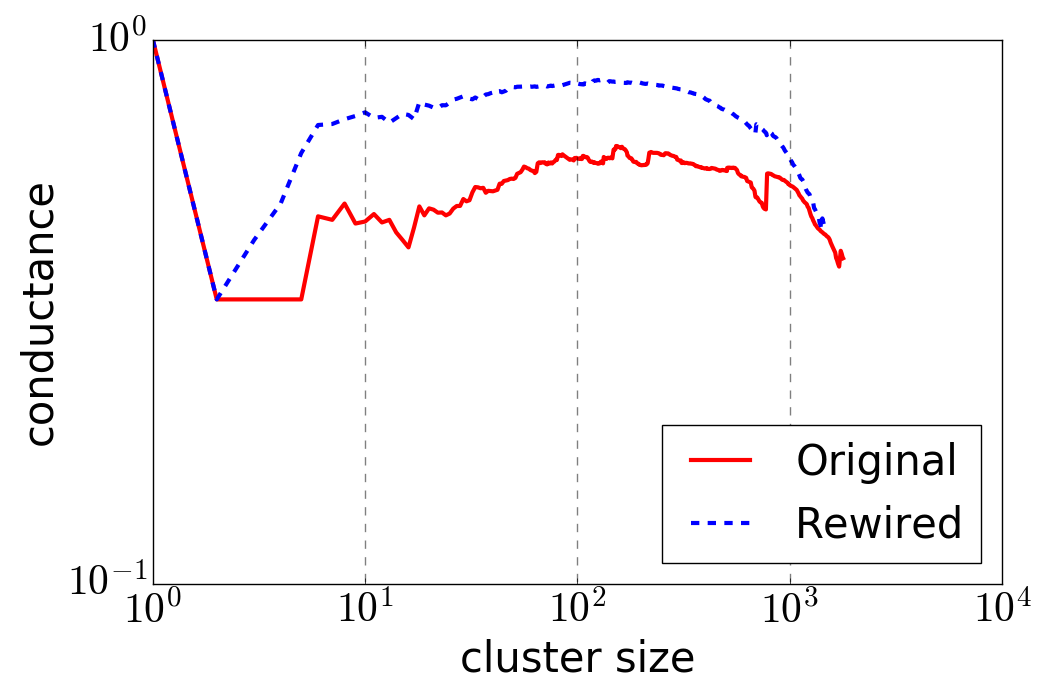}
  \label{fig:ppi_ncp}
   } \subfigure[BlogCatalog]{
       \includegraphics[width =
       0.45\columnwidth]{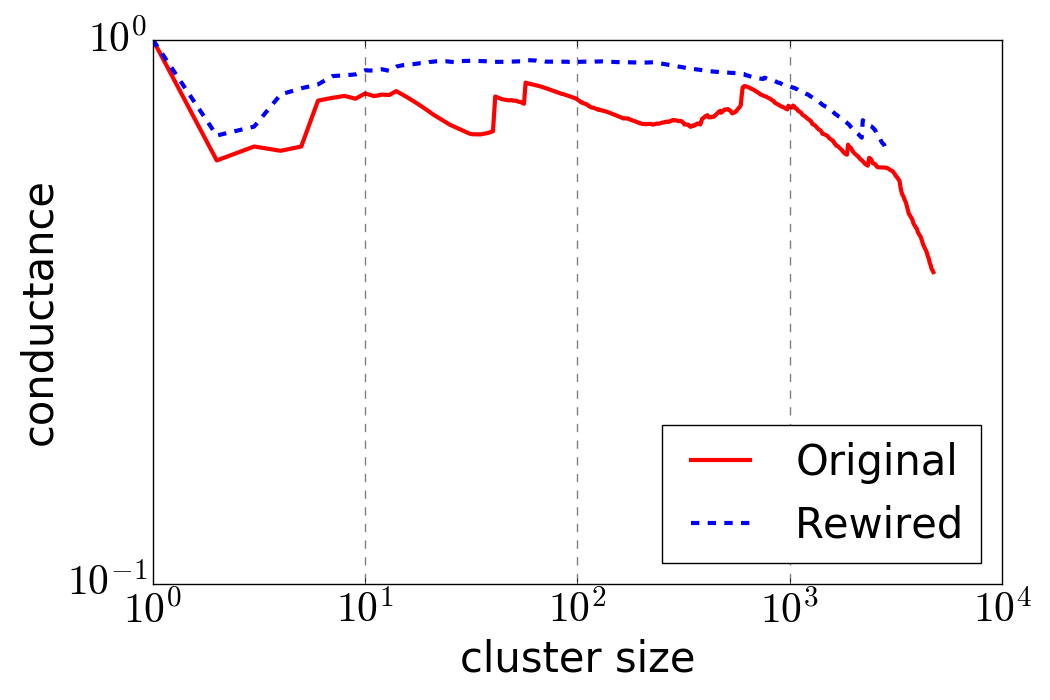}
       \label{fig:blogcatalog_ncp}
   }
 \subfigure[IMDb Germany]{
       \includegraphics[width =
       0.45\columnwidth]{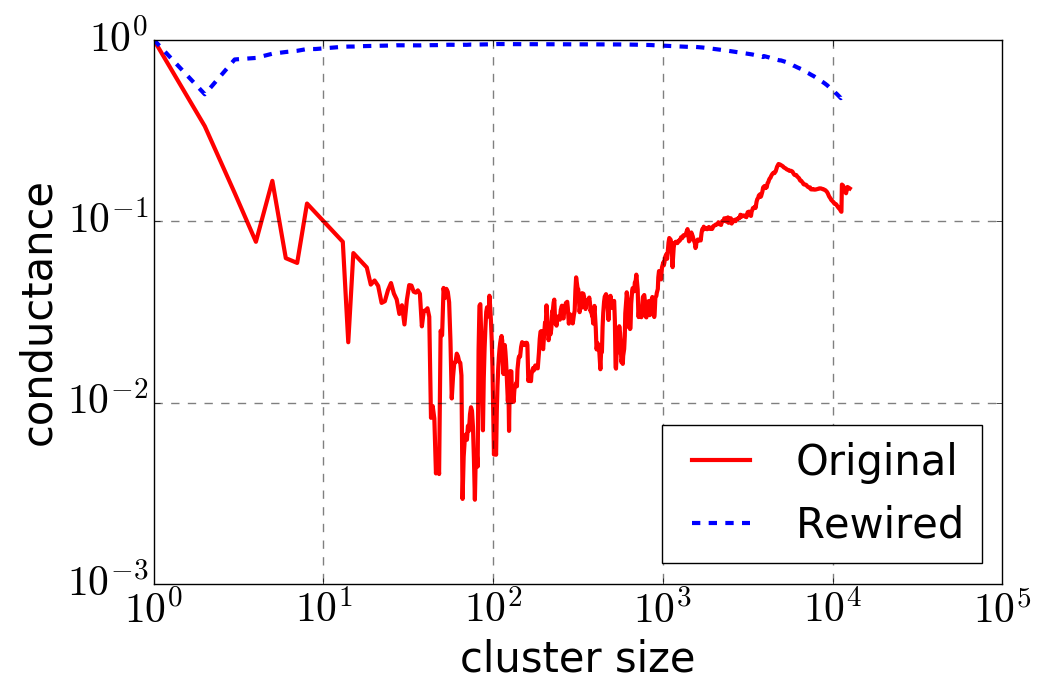}
  \label{fig:imdb_ncp}
   } \subfigure[Flickr]{
       \includegraphics[width =
       0.45\columnwidth]{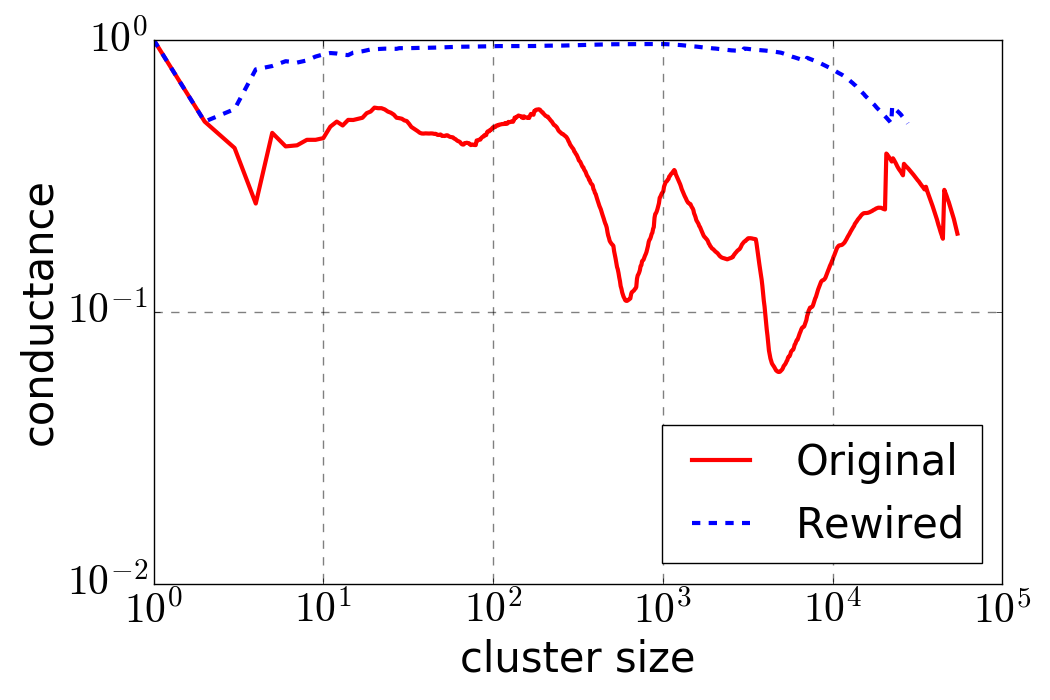}
       \label{fig:flickr_ncp}
   } %
    \caption{NCP plots for used datasets. Red, solid lines sketch the community structure of the original graph.  (Down represents better cluster quality.) Blue, dashed lines plot the structure of randomly rewired networks.}%
\label{fig:ncp_plots}
\end{center}
\end{figure}

The selection of networks captures different structures, and we use Network Community Profile (NCP) plots from \cite{LLDM08_communities_CONF,leskovec2009community,LLM10_communities_CONF,jeub2015think} to analyze them. 
The NCP depicts the best ``score" for different clusters in the graph as a function of their size. The cluster ``score'' is defined by \textit{conductance}, i.e., the ratio of edges going out of a cluster to cluster internal edges. As can be seen in Figure~\ref{fig:ncp_plots}, the IMDb Germany network has quite clear clusters of about 50 to 100 nodes. For each outgoing edge in the small clusters with near-minimum conductance value, there are about 800 internal edges. 
The three other datasets are not well separable.%
\footnote{In particular, the cluster quality is only slightly better than that of a randomly-rewired graph; \textsc{Lasagne} does particularly well for these graphs.} 
The best cluster in the Flickr graph has a size of about 5000 nodes and only about 50 internal edges for each outgoing edge.

\begin{figure}[t]
\begin{center}
    \subfigure[PPI]{
        \includegraphics[width =
        0.38\columnwidth]{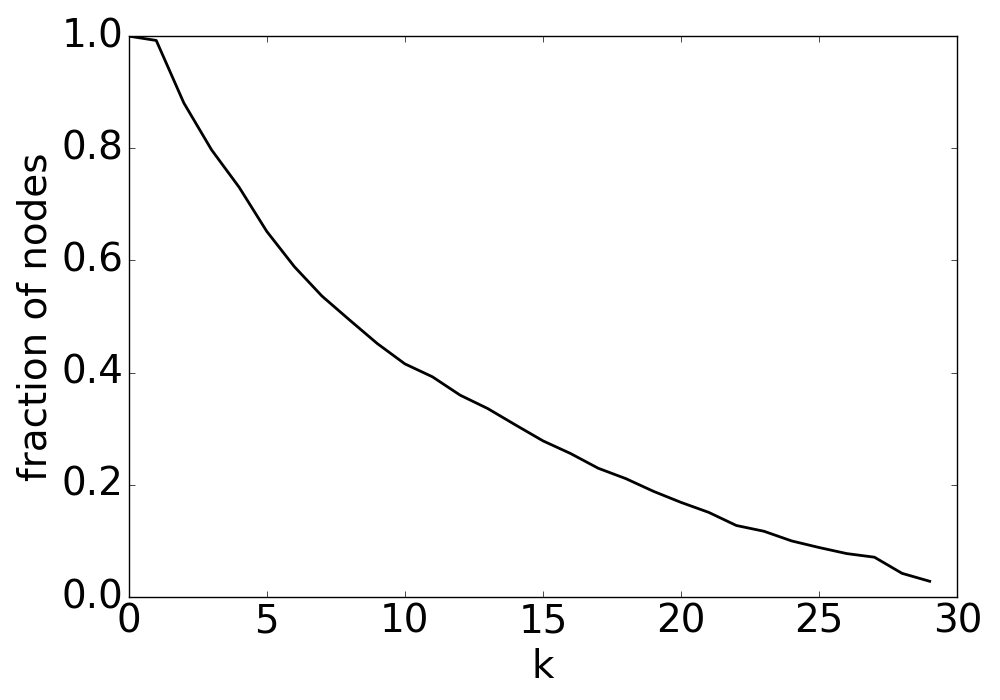}
  \label{fig:ppi_kcore}
   } \subfigure[BlogCatalog]{
       \includegraphics[width =
       0.38\columnwidth]{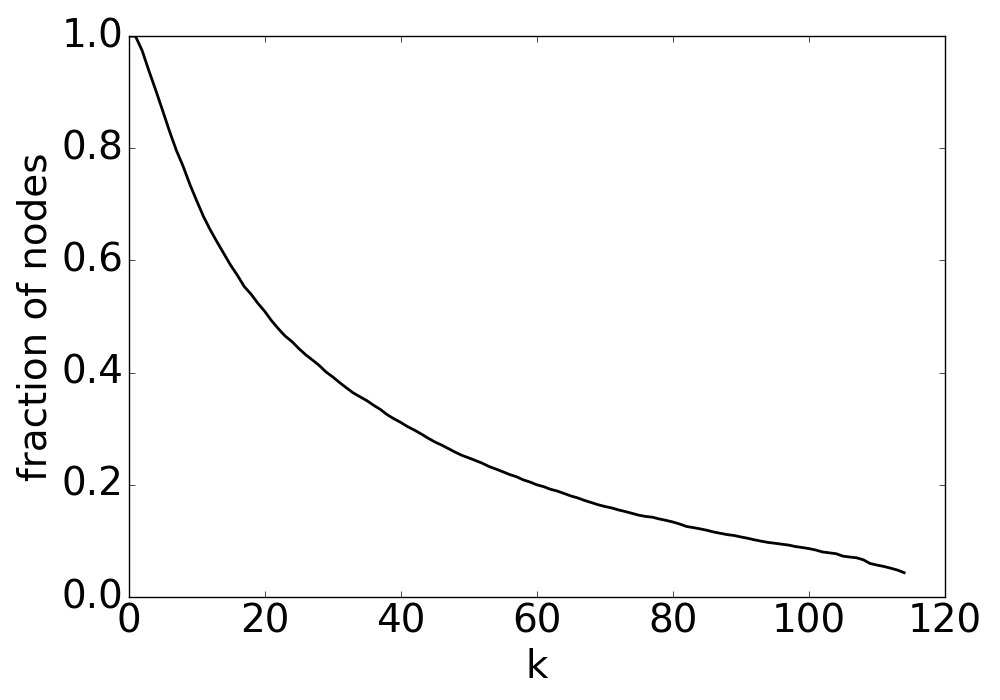}
       \label{fig:blogcatalog_kcore}
   }
   \\
 \subfigure[IMDb Germany]{
       \includegraphics[width =
       0.38\columnwidth]{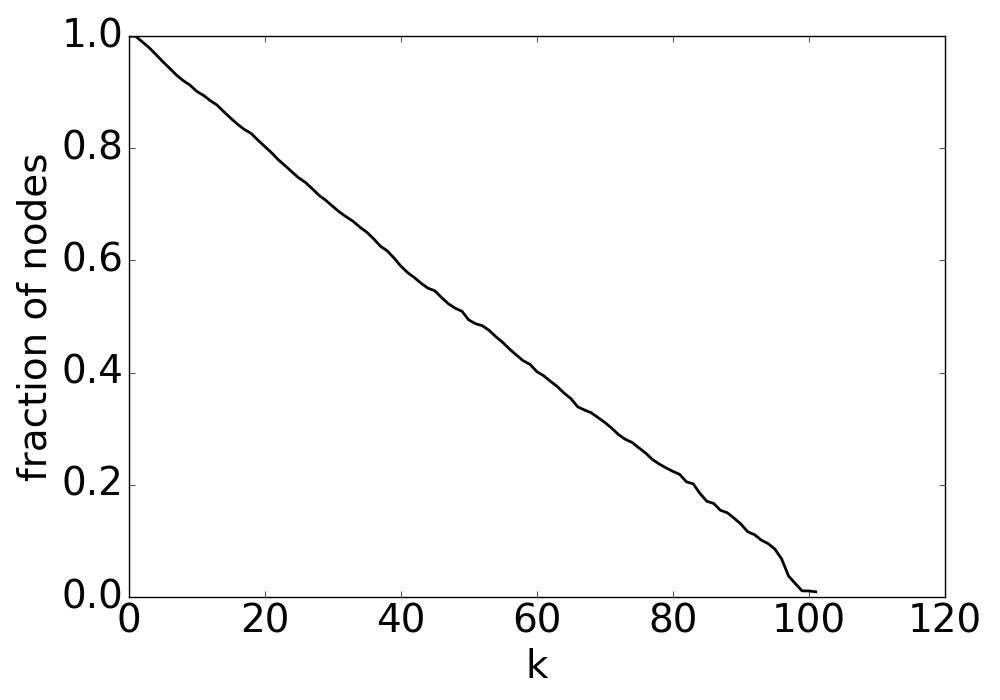}
  \label{fig:imdb_kcore}
   } \subfigure[Flickr]{
       \includegraphics[width =
       0.38\columnwidth]{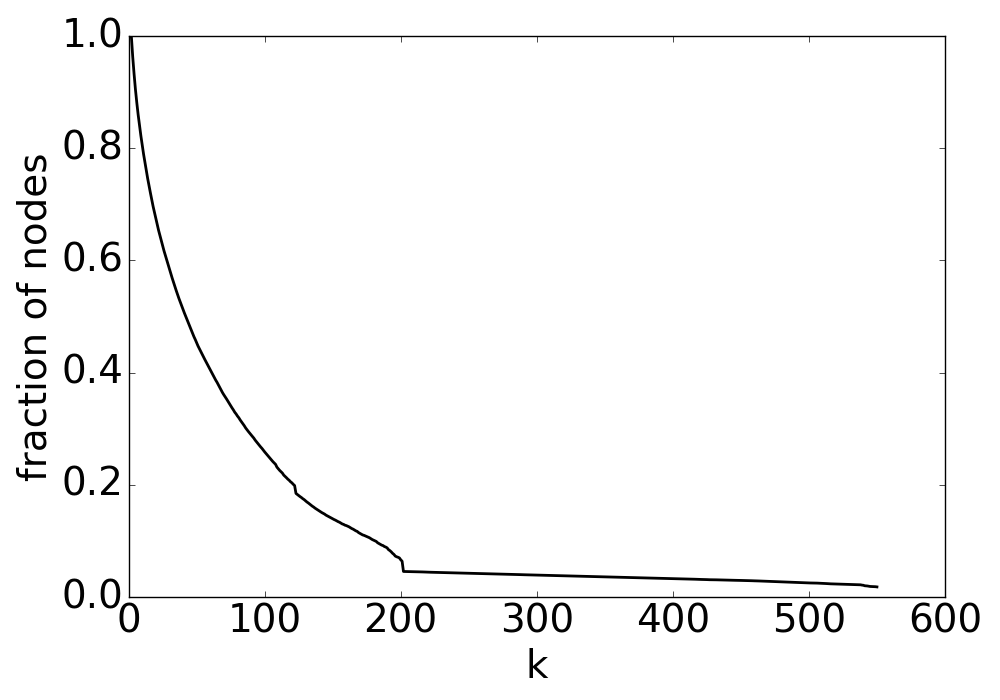}
       \label{fig:flickr_kcore}
   } %
    \caption{$k$-core plots for used datasets. Note the different scaling on the x-axes.}\label{fig:kcore_plots}%
\end{center}
\end{figure}

Following~\cite{ASM13,ASM16_IM}, we also use $k$-core information to analyze graph properties. 
The $k$-core of a graph $G$ is the maximal induced 
subgraph $H \subseteq G$ such that every node in $H$ has a degree of at least $k$. Figure~\ref{fig:kcore_plots} shows size of $k$-cores for all $k$ for all four datasets. We call a core ``deep'' if the corresponding $k$ is high. In Section \ref{sec:discussion} we discuss how size and depth of the $k$-cores affects the performance of different methods.

\subsection{Experimental Setup}
Like previous works, we use multi-label classification to evaluate the quality of the node embeddings. However, as discussed in the following, we think that the evaluation method for node representations used in \cite{perozzi2014deepwalk,tang2015line,grover2016node2vec} has a major drawback: it is hardly applicable in real world scenarios. Thus, we propose a new method for evaluating node embeddings that also relies on multi-label classification but is far closer to a real-life application scenario than the former method. 
We evaluate \textsc{Lasagne} according to both evaluation metrics.
\subsubsection{Previous Method}
Perozzi et al. \cite{perozzi2014deepwalk} made the currently used evaluation method for graph node embeddings publicly available\footnote[1]{https://github.com/phanein/deepwalk - last accessed: 2017-01-03}. The procedure is as follows: a portion of the labeled vertices is sampled randomly and used as training instances, while the remaining nodes are used for testing. The sampling approach does not preserve the percentage of samples for each class, resp. labels. After sampling, one classifier is trained for each class by using one-vs-rest logistic regression and the labels for the test instances are predicted. For the actual prediction task, this method makes recourse to information that is typically unknown. Precisely, this method uses the actual number of labels $k$ each test instance has. By sorting the predicted class probabilities and choosing the classes associated with the top $k$ probabilities, prior knowledge is incorporated into the prediction task. In real world applications, it is fairly uncommon that users have such knowledge in advance. 
A label is considered as a positive if it is among the top $k$ predicted labels, regardless its real probability value. The entire evaluation procedure is repeated 10 times and finally the average macro-$F_1$ and micro-$F_1$ scores are calculated.
\subsubsection{MoreRealisticMethod}
We propose the following modified evaluation method \textsc{MoreRealisticMethod} that reflects better the real world classification scenario where no a priori knowledge is given. Generally, we also train logistic classifiers to predict the labels of the test instances. In contrast to the method in \cite{perozzi2014deepwalk,tang2015line,grover2016node2vec}, we suggest to use a 10-fold stratified cross-validation for each one-vs-rest classifier. Using such stratified sampling is a common way to split the data into training and test set by coincidently preserving the ratio of subpopulations within the data. In this way, the prediction accuracy does not suffer from classes that may not appear in either the training or the test set due to small numbers of positive examples. 
Furthermore, we get rid of using prior knowledge to determine the positive predicted labels. Instead of ranking the probabilities and taking the labels corresponding to the top $k$ probabilities, we make the decision of labeling the test instance based on the label probabilities directly, i.e., if the probability of a label $l$ is at least 50\% we consider $l$ as positive. 

We use micro-$F_1$ and macro-$F_1$ as evaluation metrics.
Macro-$F_1$ scores build the unweighted average of $F_1$ scores for positive classes over all classifiers.
Micro-$F_1$ scores build the global average based on prescision and recall by treating each test example equally. 
We primarily focus the discussion on the macro-$F_1$ metric, but we also report the micro-$F_1$ scores. 

\subsection{Results of the MoreRealisticMethod}
The results reported in this section were obtained by using the parameter settings suggested in \cite{perozzi2014deepwalk}. We use $\gamma = 80$ as the length for the random walks performed during the \textit{DeepWalk} and \textit{node2vec} procedures.%
\footnote{If diameter $D=5,6,8,11$, then walk length $\gamma = 80$ is quite long.}
The number of random walks is $|V| \cdot r$, with $|V|$ being the number of vertices and $r=10$ being the number of random walks starting from each node in the graph. The size of the window which slides over each random walk sequence extends to at most $w=10$ in each direction of the currently regarded vertex and the dimensionality of the node embeddings is set to $d=128$. To get a fair comparison between our method and the random walk based methods, it is crucial to use similarly sized training sets for the learning procedure since larger training sets typically tend to result in higher prediction accuracy for the test phase. Thus we sample 
$$\vert T \vert = \vert V \vert \cdot \left[ \gamma \cdot r \cdot 2 \cdot \mathds{E}(\mathcal{U}(1,w)) - 2 \cdot \sum_{i=1}^w \mathds{E}(\mathcal{U}(1,i)) \right]$$
\noindent training examples which corresponds to the expected number of training instances generated by the random walk approaches. The notation $\mathds{E}(\mathcal{U}(x,y))$ denotes the expected value of a uniform distribution $\mathcal{U}$ in the interval $[x,y]$.

For \textit{node2vec} we follow the suggestions of the authors and perform full grid searches over the set $\lbrace 0.25, 0.5, 1.0, 2.0 ,4.0 \rbrace$ for both hyperparameters. The \textit{GraRep} hyperparameter $k$ is ranged from 1 to 6.
For \textsc{Lasagne} we used $\sigma=\num{0.0001}$ as significance threshold for probability updates in all empirical evaluations. We show results for different values of teleportation parameter $\alpha$.
For all datasets and all approaches we demonstrate the results when we used 90\% of the data for training and the remaining data as test set for the classification tasks.
The distributions of resulting macro $F1$ scores are visualized as box plots.
We adapted the computation of the \textit{Approximated Personalized PageRank} implemented in the \textit{Ligra} framework \cite{shun2013ligra} for our implementation. The learning procedure for the embeddings is implemented in \textit{TensorFlow} \cite{abadi2016tensorflow}. %

\begin{figure}[htb]
\begin{center}
    \subfigure[PPI]{
        \includegraphics[width =
        0.45\textwidth]{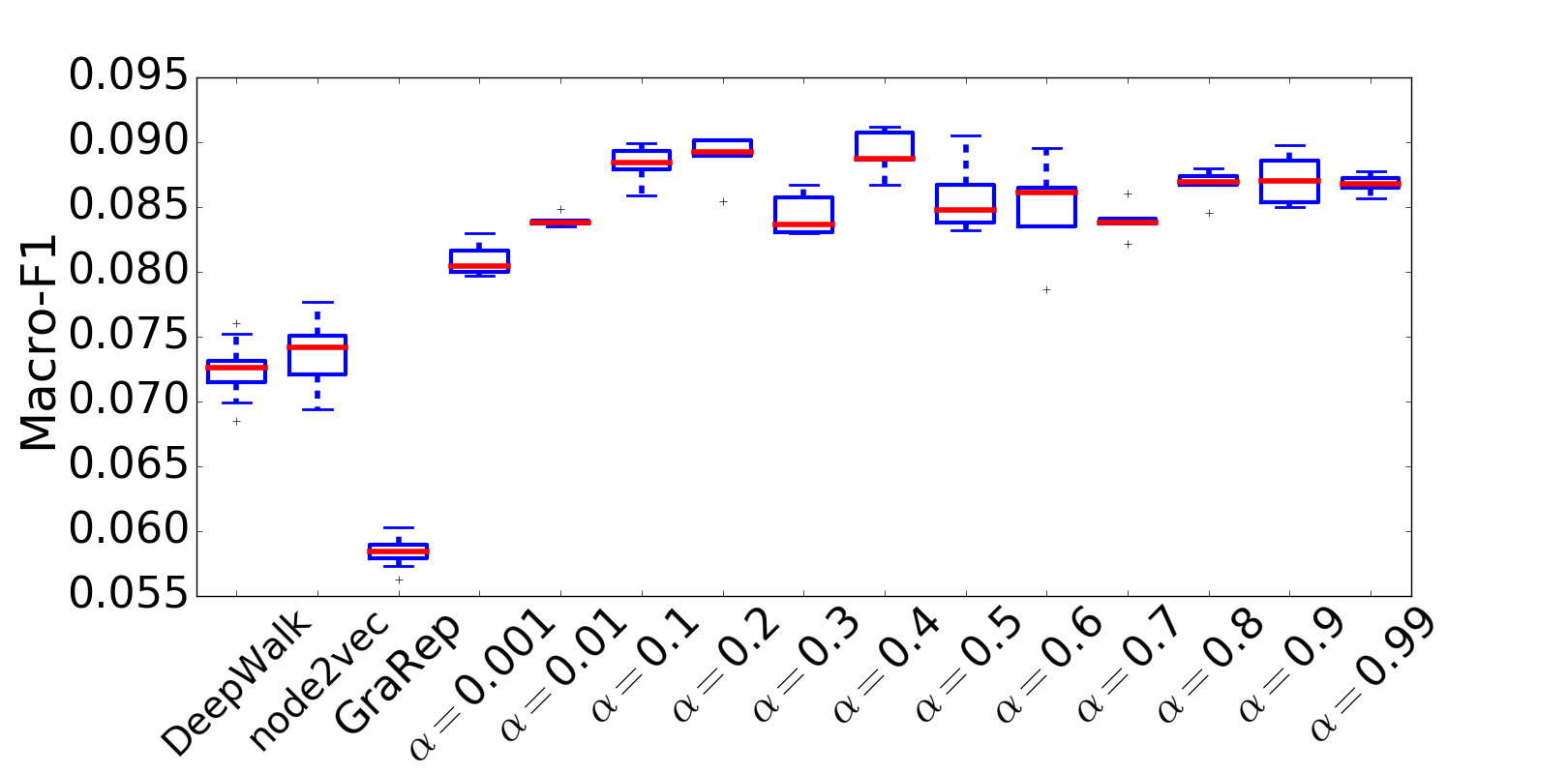}
  \label{fig:ppi_macro}
   } \subfigure[BlogCatalog]{
       \includegraphics[width =
       0.45\textwidth]{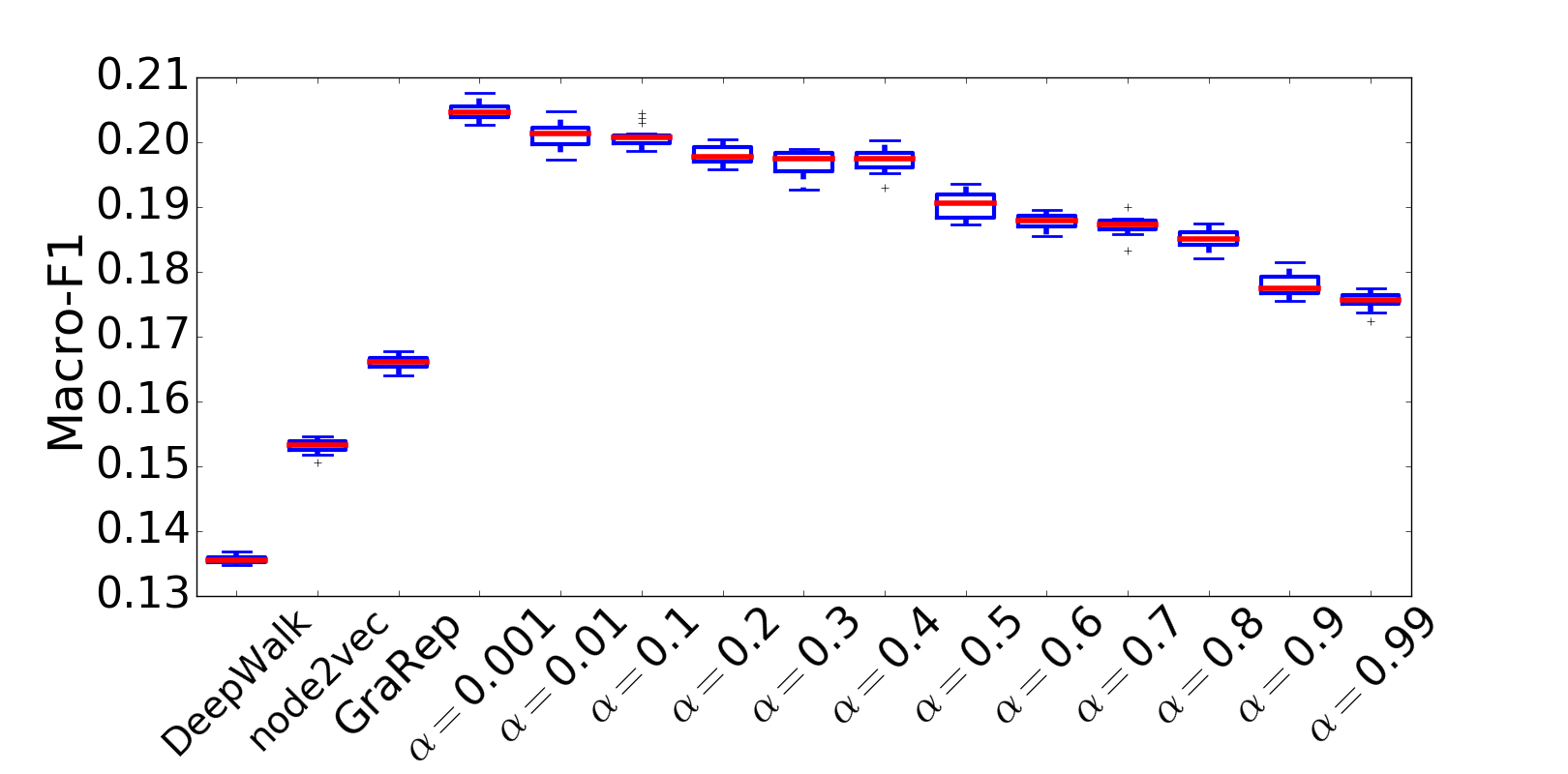}
       \label{fig:blogcatalog_macro}
   }
 \subfigure[IMDb Germany]{
       \includegraphics[width =
       0.45\textwidth]{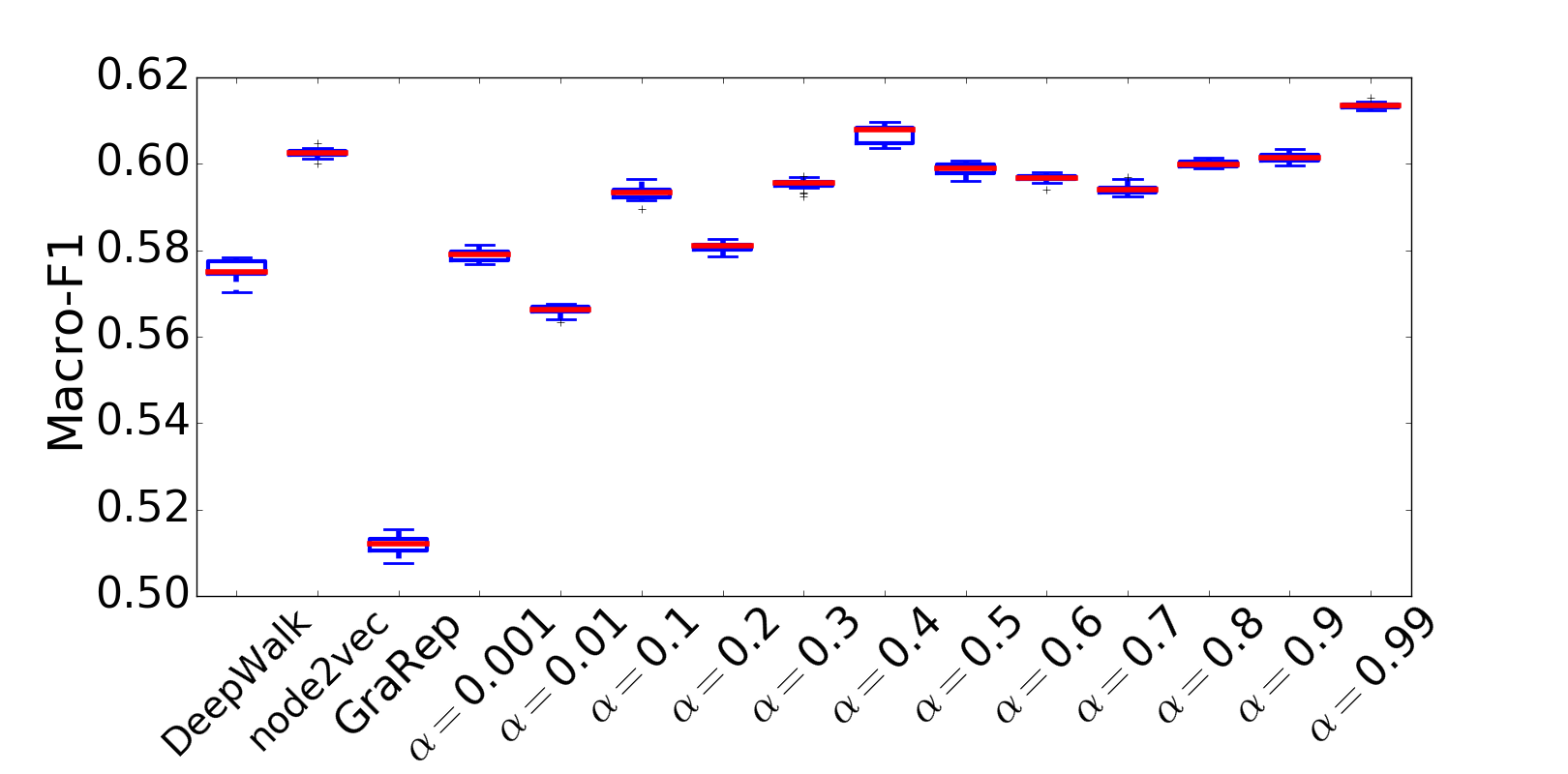}
  \label{fig:imdb_germany_macro}
   } \subfigure[Flickr]{
       \includegraphics[width =
       0.45\textwidth]{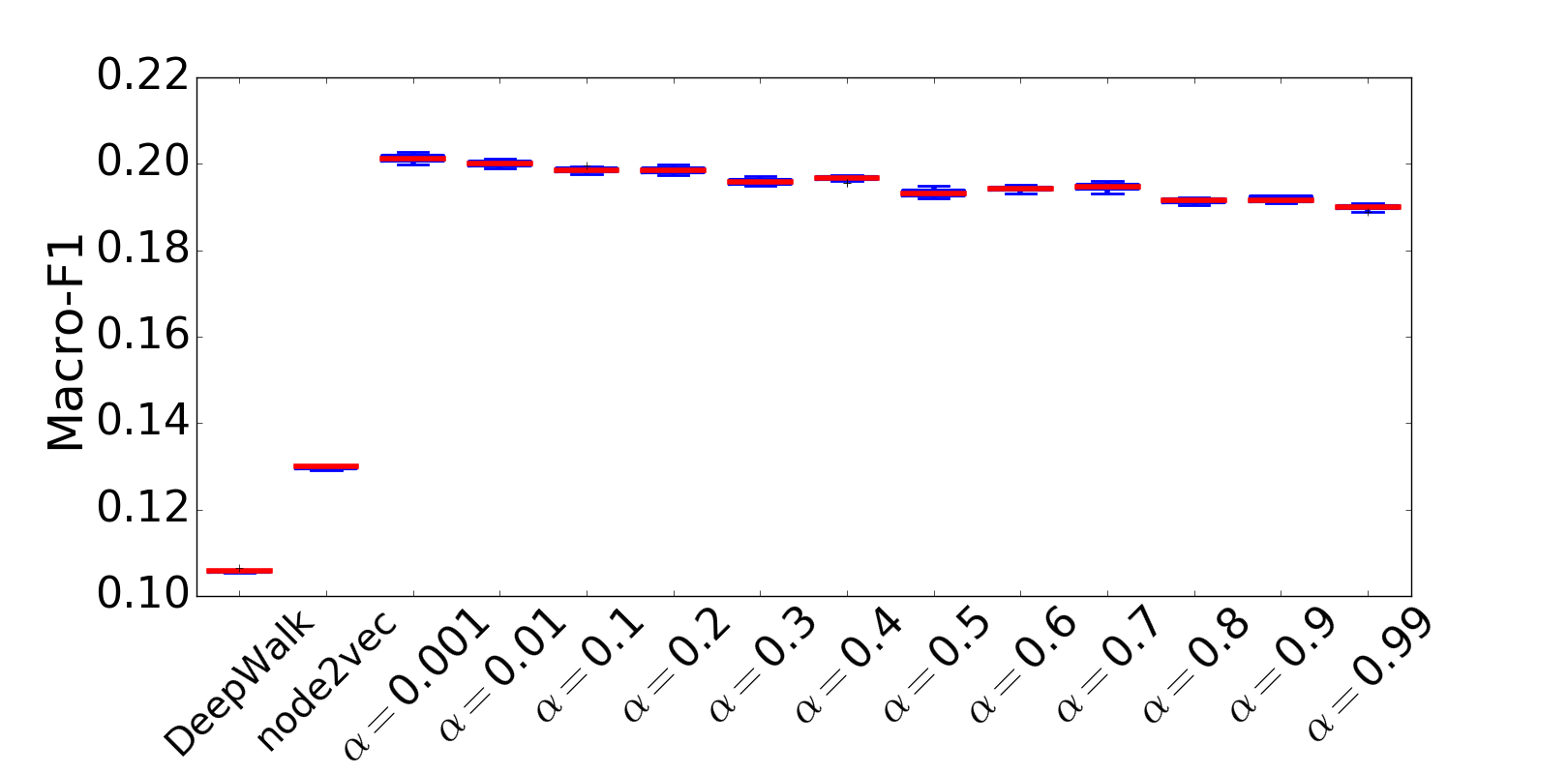}
       \label{fig:flickr_macro}
   } %
    \caption{Macro-$F_1$ scores achieved by doing multi-label classification as downstream task for the considered representation learning techniques. \textsc{Lasagne} scores are presented for different values of parameter $\alpha$.}%
    \label{fig:macroscores}
\end{center}
\end{figure}

Figure~\ref{fig:macroscores} shows the macro-$F_1$ scores for all methods and datasets when applying embeddings for multi-label classification. \textsc{Lasagne} overcomes the competitors for each dataset. 

For the PPI network, c.f. Figure~\ref{fig:ppi_macro}, the scores are steadily over 8\% for all $\alpha$ values, while the random walk approaches reach scores between 7\% and 7.5\%. The best \textit{node2vec} setting is $p=4$ and $q=1$, which corresponds to a rather low willingness to allow the random walks to return to already visited nodes. This meets the outcomes of \textsc{Lasagne}, which are best for small $\alpha$ values. The generally low prediction quality for all approaches, and especially the bad score for \textit{GraRep}, may indicate that the distribution of class labels do not follow any representative, local patterns and hence are hardly graspable within local structures (at least in this set of data). 
The results for BlogCatalog are even more clear. \textsc{Lasagne} improves the best competitor by approximately 23\%. As can be seen in Figure~\ref{fig:blogcatalog_macro} the performance of \textsc{Lasagne} decreases almost monotonically with increasing values for $\alpha$. This means that the neighbors which describe a node best are not extremely local. The best \textit{node2vec} setting, i.e., $p=0.25$ and $q=0.25$, confirms this results. Recalling Figure 2 from \cite{grover2016node2vec}, the 2nd order random walks are biased towards leaving the neighborhoods.
For IMDb Germany, c.f. Figure~\ref{fig:imdb_germany_macro}, the best result of \textsc{Lasagne}, which is for $\alpha=0.99$, is only slighty better than the best results achieved with \textit{node2vec}. Since \textsc{Lasagne} is, as well as \textit{node2vec} with parameter setting $p=0.25, q=4$, able to stay extremely local, both approaches reach high prediction scores on this dataset where the labels are concentrated in low conductance clusters.
Using the Flickr network, \textsc{Lasagne} reaches the highest improvement over the other random walk based methods, i.e. more than 33\%. The results behave similar to the ones for the BlogCatalog data, but in contrast the scores remain more stable. Indeed, the drop between the smallest and largest selected $\alpha$ values is only 1\%.
As mentioned previously, we could not run \textit{GraRep} on Flickr, because of its size. 

\begin{figure}[htb]
\begin{center}
    \subfigure[PPI]{
        \includegraphics[width =
        0.45\textwidth]{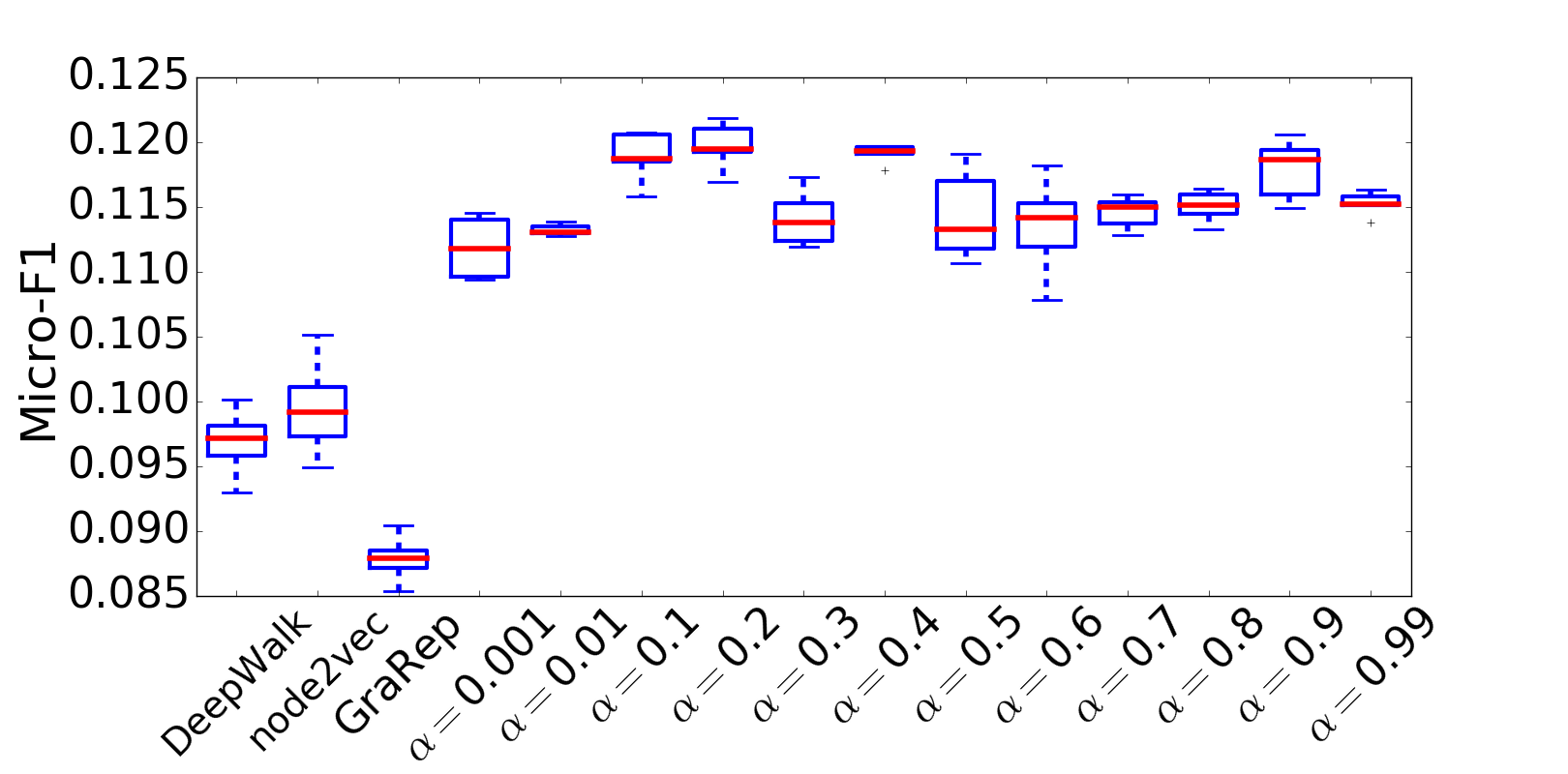}
  \label{fig:ppi_micro}
   } \subfigure[BlogCatalog]{
       \includegraphics[width =
       0.45\textwidth]{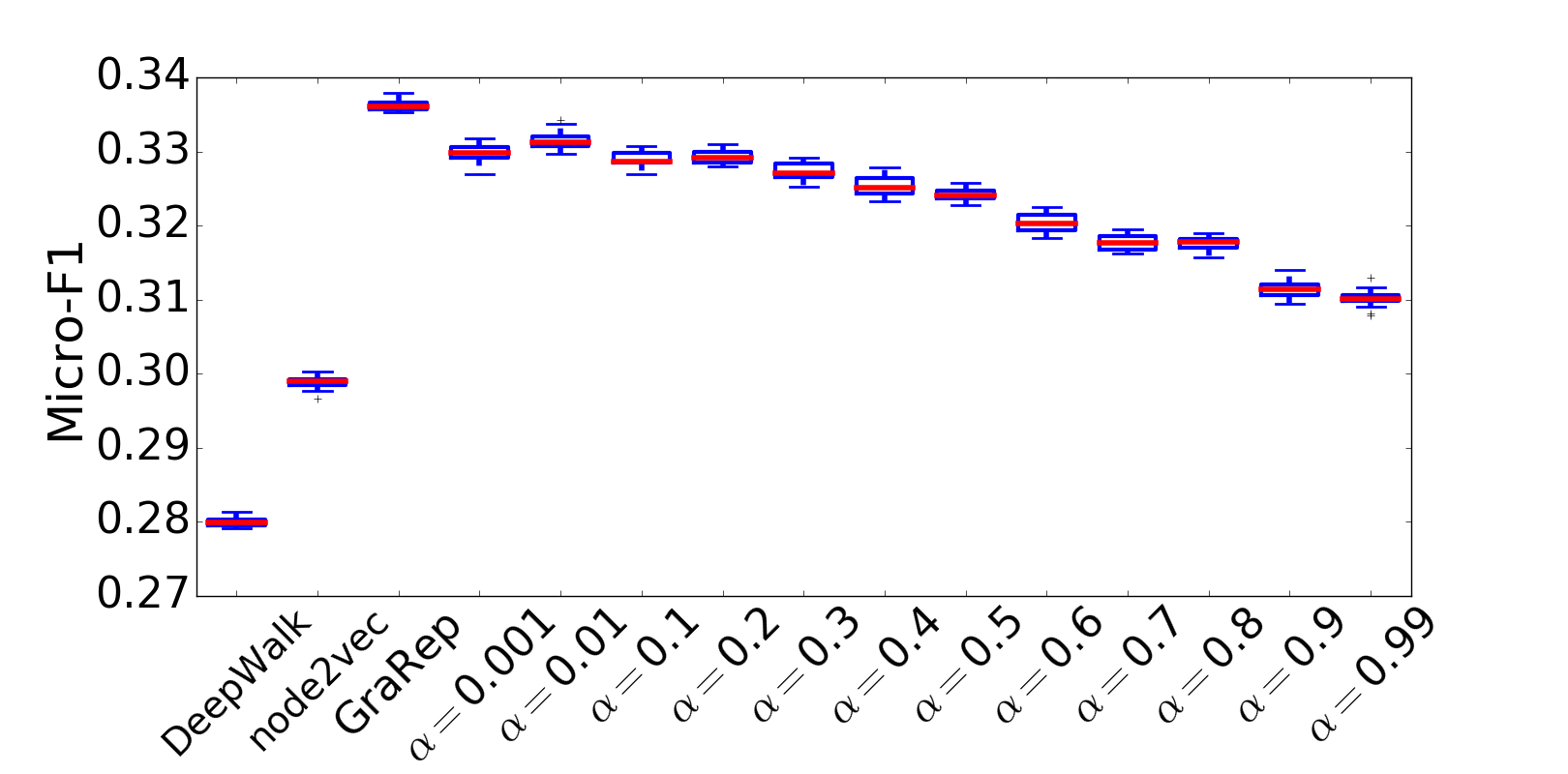}
       \label{fig:blogcatalog_micro}
   }
 \subfigure[IMDb Germany]{
       \includegraphics[width =
       0.45\textwidth]{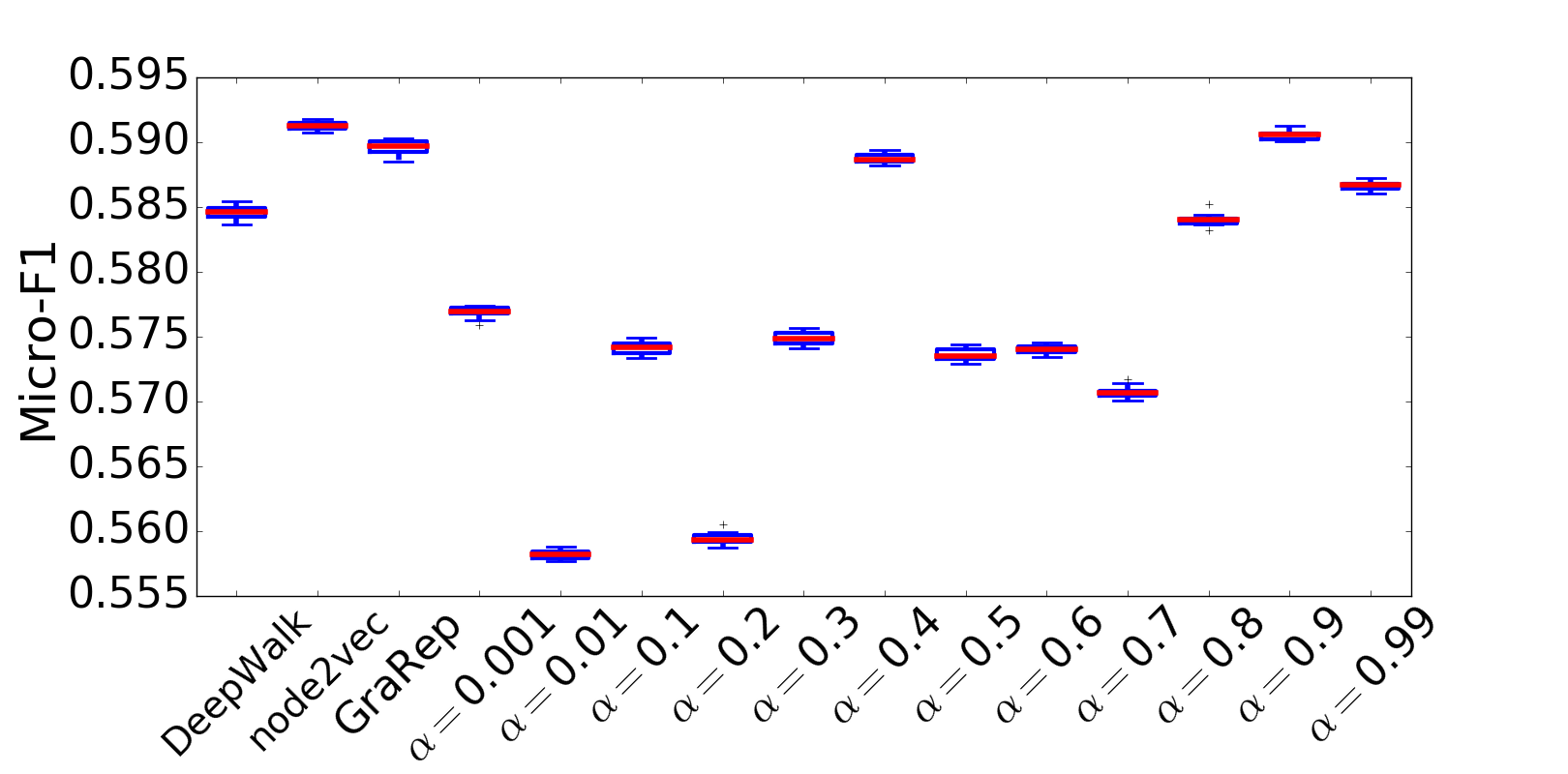}
  \label{fig:imdb_micro}
   } \subfigure[Flickr]{
       \includegraphics[width =
       0.45\textwidth]{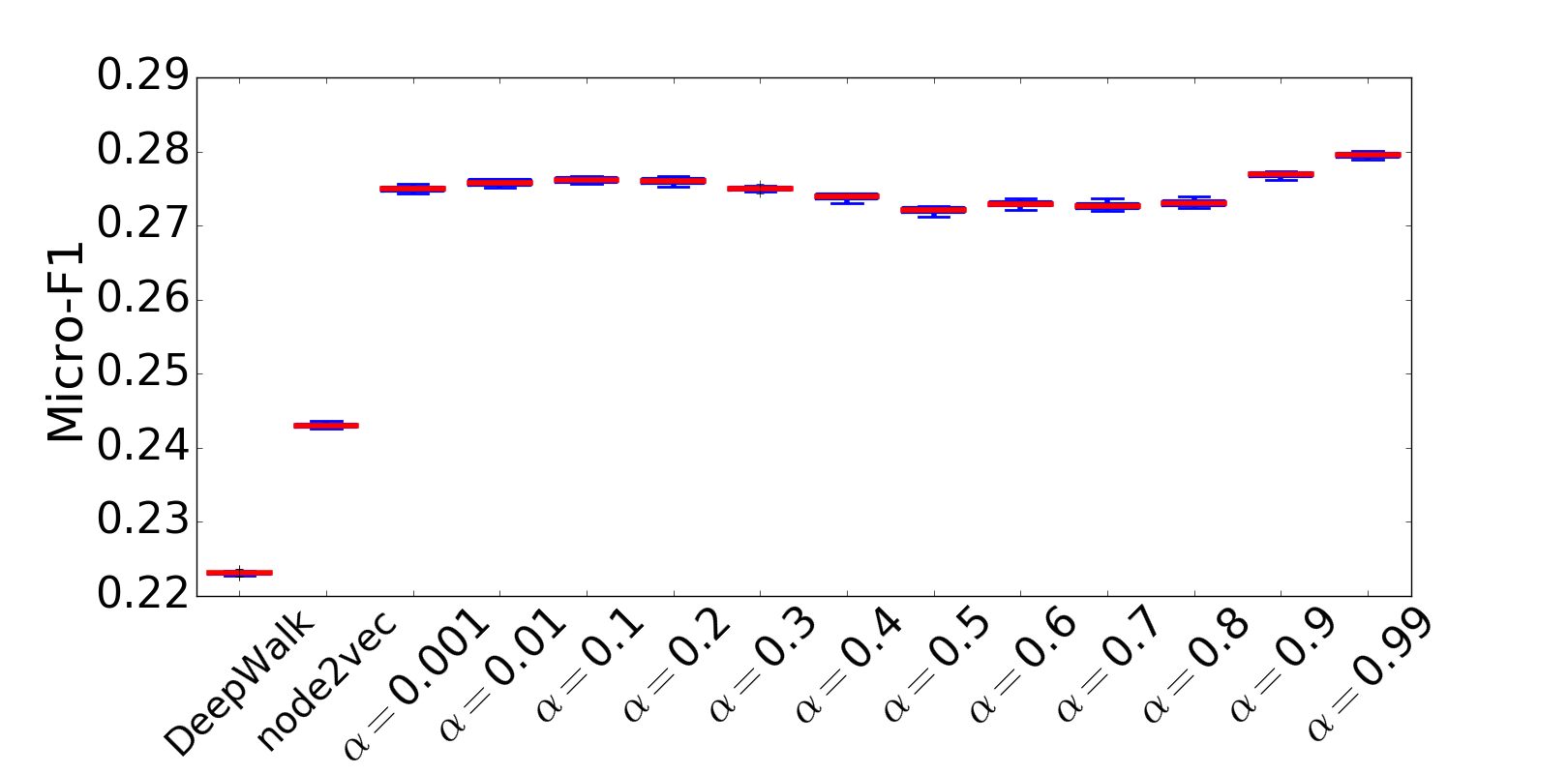}
       \label{fig:flickr_micro}
   } %
    \caption{Micro-$F_1$ scores achieved by doing multi-label classification as downstream task for the considered representation learning techniques. \textsc{Lasagne} scores are presented for different values of parameter $\alpha$.}%
    \label{fig:microscores}
\end{center}
\end{figure}

Figure~\ref{fig:microscores} shows the micro-$F_1$ scores achieved with the same settings as used for the macro-$F_1$ score evaluation. The results show that the micro scores are higher than the macro scores for all datasets except for IMDb Germany. Also the relative differences between the results for \textsc{Lasagne} and the best competitor are higher for the macro-$F_1$ scores than for the micro-$F_1$ scores. This is due to the micro score metric effectively gives higher weight to larger classes. 
This may be justified by the results depicted in Figure~\ref{fig:all_single_classes} (discussed below). 
Since \textsc{Lasagne} performs better for smaller classes which, except for IMDb Germany, are the vast majority of classes, the macro-$F_1$ scores take benefit due to weighting each class equally independent from the class sizes. 
Recalling that the micro-$F_1$ considers the sizes of the classes, the performance improvements for this score are reasoned by the fact that \textsc{Lasagne} performs better on smaller classes and similarly good to random walk based methodologies on larger classes.

An important summary point from Figures~\ref{fig:macroscores} and~\ref{fig:microscores} is that, in the case of graphs without even small-sized good conductance clusters, the performance of \textsc{Lasagne} clearly overcomes the performance from random walk based methods. 
On the other hand, for graphs that have an upward-sloping NCP and thus small-sized good conductance clusters, \textsc{Lasagne} shows similar prediction quality to random walk based methods.
In particular, while we are never worse than previous methods, we observe the weakest improvement for IMDb Germany, which is consistent with Figure~\ref{fig:imdb_ncp}, where the upward-sloping NCP suggests relatively good local structure, and we observe the strongest improvement for the Flickr network, which is consistent with Figures~\ref{fig:flickr_ncp} and~\ref{fig:flickr_kcore}, which indicate a relatively flat NCP  and many deep $k$-core nodes.

\subsection{Results of the Former Evaluation Method}

Tables~\ref{tab:n2v_macro} and \ref{tab:n2v_micro} show the macro-$F_1$ scores, resp. the micro-$F_1$ scores when applying the evaluation proposed by \cite{perozzi2014deepwalk} and using 90\% of the node representations for training. While \textit{GraRep} shows the best results on PPI, the performance of the \textsc{Lasagne} embeddings clearly overcomes the competitors when testing on the considered social networks, similar to the results in our more realistic (and more refined) evaluation. 

\begin{table}[t]  %
\begin{center}
  \begin{tabular}{ | c | c | c | c | c | }
    \hline
    \multirow{2}{*}{Algorithm} & \multicolumn{4}{c |}{Dataset} \\ \cline{2-5}
     & PPI & BlogCatalog & IMDb Ger & Flickr \\ \hline
    DeepWalk & 0.1747 & 0.2221 & 0.6868 & 0.2104 \\ \hline
    node2vec & 0.1930 & 0.2418 & 0.6996 & 0.2349 \\ \hline
    GraRep & \textbf{0.1991} & 0.2231 & 0.5770 & - \\ \hline
    \textsc{Lasagne} & 0.1835 & \textbf{0.2843} & \textbf{0.7042} & \textbf{0.2930} \\ \hline
  \end{tabular}
  \end{center}
  \caption{Macro-F$_1$ scores for multi-label classification when using former evaluation method and 90\% of instances for training.}
  \label{tab:n2v_macro}
\end{table}

\begin{table}[t]  %
\begin{center}
  \begin{tabular}{ | c | c | c | c | c | }
    \hline
    \multirow{2}{*}{Algorithm} & \multicolumn{4}{c |}{Dataset} \\ \cline{2-5}
     & PPI & BlogCatalog & IMDb Ger & Flickr \\ \hline
    DeepWalk & 0.2206 & 0.3889 & 0.7043 & 0.3762 \\ \hline
    node2vec & 0.2293 & 0.3963 & \textbf{0.7060} & 0.3841 \\ \hline
    GraRep & \textbf{0.2487} & 0.3913 & 0.6648 & - \\ \hline
    \textsc{Lasagne} & 0.2216 & \textbf{0.4116} & 0.6967 & \textbf{0.4078} \\ \hline
  \end{tabular}
  \end{center}
  \caption{Micro-F$_1$ scores for multi-label classification when using former evaluation method and 90\% of instances for training.} \label{tab:n2v_micro}
\end{table}

\subsection{Link Prediction}
For completeness, Table~\ref{tab:link_prediction} reports the results when applying the embeddings retrieved by \textsc{Lasagne} on the link prediciton task. The experimental setup is borrowed from \cite{grover2016node2vec} which means that we removed 50\% of the edges of each graph, learned the representations on the remaining graph and finally predict the existence of the removed edges by using a binary classifier. The classifier is trained with the remaining 50\% of edges as positive examples and the same amount of non-existent edges as negative samples. The edges were embedded by using one of the embedding methods documented in Table~\ref{tab:link_prediction}. Hence, an edge embedding is the combination of the representation of the nodes joined by the corresponding edge according to the specified method. As evaluation metric we also use the well-known Area Under the Curve (AUC) score. %
For \textsc{Lasagne} and \textit{node2vec} we used the same set of parameter settings as for multi-label classification. The reported results are the best results that were achieved by all settings.
We consider the following graphs for link prediction: %
\begin{itemize}
\item Facebook \cite{snapnets}: This is a social network consisting of friend lists from facebook. The network consists of 4,039 nodes that represent users and 88,234 edges which represent friendships between the corresponding users. 
\item BlogCatalog \cite{tang2009relational}: This is the same network as in Table~\ref{tab:statistics}.  %
\item arXiv Astro-Ph \cite{snapnets}: This is a collaboration network which covers scientific collaborations. It consists of 18,772 nodes and 198,110 edges. Each node represents an author and edges connect authors who collaborated on a joint work submitted to arXiv astro physics category. 
\end{itemize}
Facebook (which  has a relatively flat NCP~\cite{leskovec2009community,jeub2015think}) and arXiv Astro-Ph (which has an upward-sloping NCP~\cite{leskovec2009community,jeub2015think}) were also used in \textit{node2vec} \cite{grover2016node2vec} for link prediction.

\begin{table}[t]  %
\begin{center}
  \begin{tabular}{ | c | c | c | c | c | }
    \hline
    \multirow{2}{*}{op} & \multirow{2}{*}{Algorithm} & \multicolumn{3}{c |}{Dataset} \\ \cline{3-5}
     &  & facebook & arXiv & BlogCatalog \\ \hline
     \hline
    \multirow{4}{*}{a)} & DeepWalk & 0.7240 & 0.7002 & 0.7921 \\ \cline{2-5}
     & node2vec & 0.7223 & 0.7259 & 0.8108 \\ \cline{2-5}
     & GraRep & 0.7495 & 0.7097 & 0.8759 \\ \cline{2-5}
     & \textsc{Lasagne} & 0.7069 & 0.7195 & 0.8701 \\ \hline
     \hline
     \multirow{4}{*}{b)} & DeepWalk & 0.9610 & 0.8632 & 0.7187 \\ \cline{2-5}
     & node2vec & 0.9644 & 0.8770 & 0.7359 \\ \cline{2-5}
     & GraRep & 0.9629 & 0.7494 & 0.8846 \\ \cline{2-5}
     & \textsc{Lasagne} & 0.9628 & 0.8715 & 0.8281 \\ \hline
     \hline
    \multirow{4}{*}{c)} & DeepWalk & 0.9606 & 0.8438 & 0.7799 \\ \cline{2-5}
     & node2vec & 0.9642 & 0.8499 & 0.8044 \\ \cline{2-5}
     & GraRep & 0.9621 & 0.7980 & 0.8713 \\ \cline{2-5}
     & \textsc{Lasagne} & 0.9072 & 0.7036 & 0.7017 \\ \hline
     \hline
     \multirow{4}{*}{d)} & DeepWalk & 0.9593 & 0.8450 & 0.7844 \\ \cline{2-5}
     & node2vec & 0.9646 & 0.8523 & 0.8074 \\ \cline{2-5}
     & GraRep & 0.9635 & 0.7664 & 0.8731 \\ \cline{2-5}
     & \textsc{Lasagne} & 0.9111 & 0.7053 & 0.7045 \\ \hline
     \hline
     \multirow{4}{*}{$jac$} & DeepWalk & 0.8435 & 0.7357 & 0.5525 \\ \cline{2-5}
     & node2vec & 0.8509 & 0.7381 & 0.5644 \\ \cline{2-5}
     & GraRep & 0.8418 & 0.4980 & 0.5567 \\ \cline{2-5}
     & \textsc{Lasagne} & 0.9256 & 0.7361 & 0.5337 \\ \hline
  \end{tabular}
  \end{center}
  \caption{Results for Link Prediciton; Metric: AUC scores of predictions retrieved by binary classifiers resp. Jaccard similarity measure; Operators used for edge embedding: a) Average: $\frac{f_i(u) + f_i(v)}{2}$, b) Hadamard: $f_i(u) \cdot f_i(v)$, c) Weighted L1: $\vert f_i(u) - f_i(v)\vert$, d) Weighted L2: $\vert f_i(u) - f_i(v)\vert ^2$, with $f_i(x)$ being the $i$-th component of node $x$ \cite{grover2016node2vec}; $jac$: Jaccard similarity measure}\label{tab:link_prediction}
\end{table}

\noindent 
Overall, these results show that the \textsc{Lasagne} embeddings perform as well as the representations learned by \textit{node2vec} when considering the facebook dataset or the arXiv dataset. The actual differences between the best results are less than 1\%. For the BlogCatalog data, the representations retrieved by \textsc{Lasagne} even improve the best prediction score reached by the random walk based competitors.
Disregarding the edge embedding methods proposed in \cite{grover2016node2vec} and using the jaccard similarity ($jac$), i.e., $jac(u,v) = \frac{\mathcal{N}_k(u) \cap \mathcal{N}_k(v)}{\mathcal{N}_k(u) \cup \mathcal{N}_k(v)}$ with $\mathcal{N}_k(u)$ being the $k$ nearest neighbors of node $u$ in the embedded space, instead, \textsc{Lasagne} also shows similar results as the competitors and yields the best results on the facebook data. 
\subsection{Explaining our improved empirical results}
\label{sec:discussion}

In this section, we present additional empirical results aimed at explaining in terms of local properties of the graph topology \textsc{Lasagne}'s improved performance.
\textsc{Lasagne} improves previous methods by considering more finely the structure of the graph around each node.
In particular, we compute local node neighborhood by touching only the \textit{relevant} neighbors of each node, which leaves the major part of the graph unconsidered.\footnote{Both our approach and the previous approaches which we improve use some sort of random walk to construct ``sentences,'' each ``word'' of which is a node from the graph, and then they call the \textit{word2vec} method.  Essentially, our two improvements use APPR to more precisely engineer in locality, thereby leading to higher-quality contexts.  The importance and sensitivity of such preprocessing is well known in natural language processing.}
For the node $a$ we call $b$ its \textit{relevant} neighbor if $b$ has  high probability to be visited by random walk with restart starting from $a$.

\subsubsection{Locality for nodes with different degrees}
Previous random walk based methods follow a similar scheme, except they simulate \emph{long} random walks in the graph. 
For each node occurring in one of the random walks, a window of dynamic size which contains nodes visited previously and after that node is used to determine the context. 
The actual extension of the window to each side is sampled  each time uniformly from the interval $[1, w]$, where $w$ is a hyperparameter (that is the same for all nodes in the graph). 
For example, while simulating the random walks the \textit{DeepWalk} \cite{perozzi2014deepwalk} algorithm selects the next node fully arbitrary among the neighbors of the last visited node. \textit{node2vec} \cite{grover2016node2vec} generally gives more control over the context selection due to its hyperparameters and thus allows the prioritization of closer resp. farther neighbors. 
This flexibility comes to the cost of an expensive preprocessing step which is quadratic in node degree.

Nevertheless, even with this expensive preprocessing, existing methods fail to adapt to the local graph structure.
When random walks are used to obtain neighbors, nodes having very low probability to be visited also appear among the considered neighbors. 
Nodes having high probabilities to be visited appear more frequently.  %
However, the cumulative probability of low probability nodes may still be significant.
The wider the window is, the more far away neighbors end up in it. 
However, smaller window sizes will not help to tackle the problem with low probability neighbors, since the nodes in sparse graph areas may have distant neighbors with high probability to be visited by random walk. Grover et al. \cite{grover2016node2vec} even show, that they achieve better results with larger window sizes. However, since the same window size is used for all nodes in the same graph, the distributions of hop distances of nodes to their neighbors are similar and barely adapt to local node neighborhood. 
\begin{figure}[t]
    \subfigure[\textit{node2vec}]{
        \includegraphics[width =
        0.46\columnwidth]{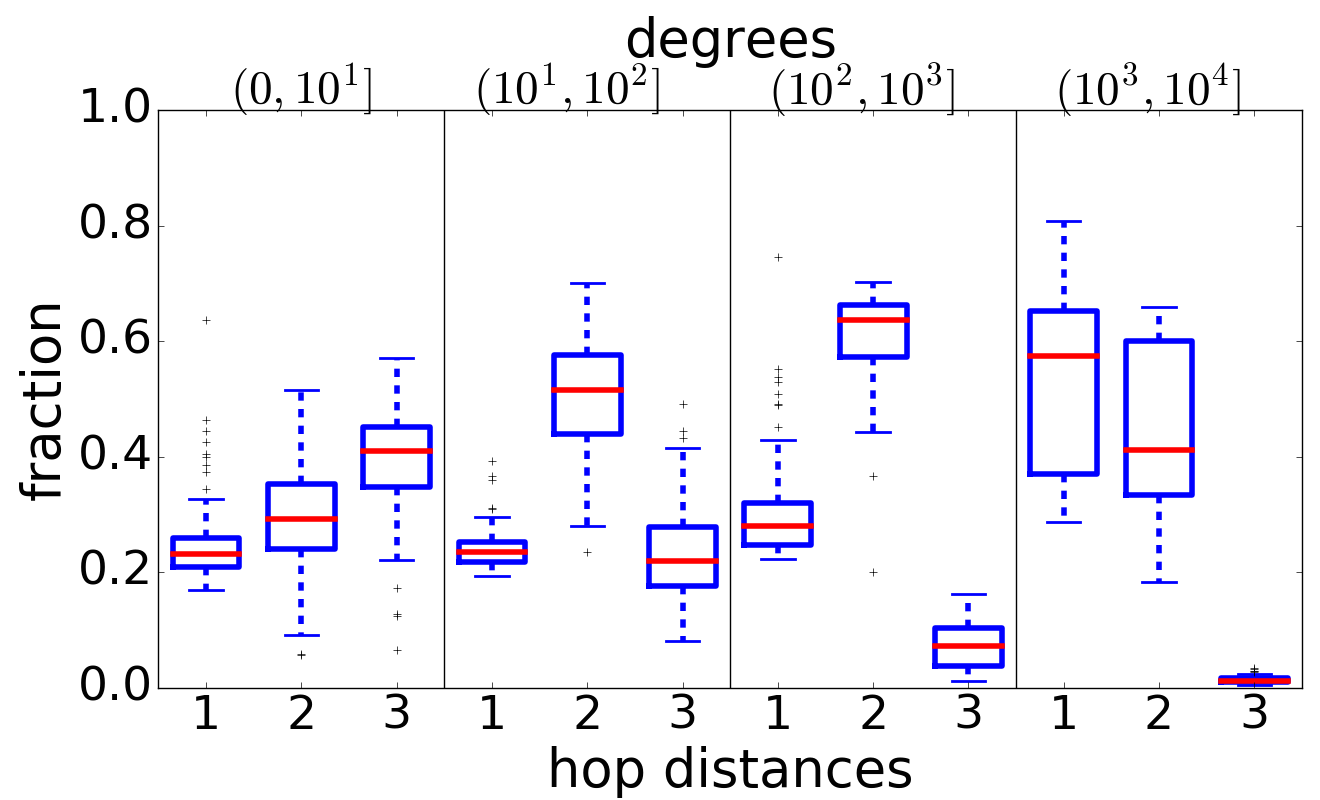}
  \label{fig:node2vec_hops_to_degree}
   } \subfigure[\textsc{Lasagne}]{
       \includegraphics[width =
       0.46\columnwidth]{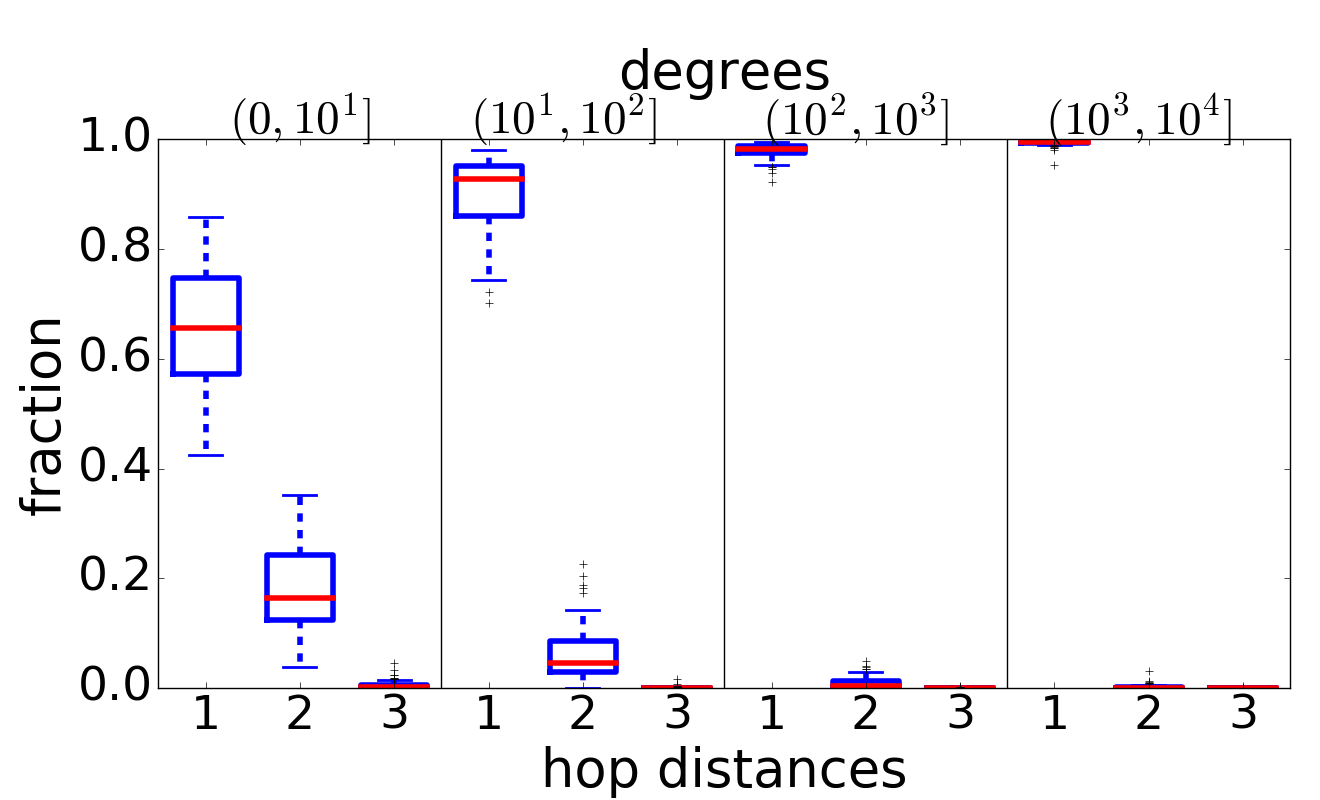}
       \label{fig:lasagne_hops_to_degree}
   }
    \caption{Distributions of hop distances to neighbors from nodes with different degrees. These plots visualize the ability to adjust to differently dense areas in the graph for \textit{node2vec} (left, not well) and \textsc{Lasagne} (right, very well).}\label{fig:distance_to_neighbors}%
\end{figure}
    
To confirm this intuition, we computed the hop distances to the nodes considered as context by \textit{node2vec} and \textit{DeepWalk} algorithms for different datasets. 
For all of them, we observed similar behavior, i.e., the level of locality was barely adapted with increasing node degree, c.f., Figure~\ref{fig:node2vec_hops_to_degree}. 
Note that the \textit{node2vec} parameters were set to $p=0.25$ and $q=4.0$, which constrains the random walks to capture very local neighborhoods (but in a non-adaptive manner).
The distributions of hop distances to the neighbors found by the \textsc{Lasagne} algorithm are very similar per dataset; an example is depicted in Figure\ref{fig:lasagne_hops_to_degree}.
In contrast to the previous methods, \textsc{Lasagne} adapts to the local node environment, i.e., for the high degree nodes only the neighbors with the highest probability to be visited by the random walk are considered as context. 
Consequently, we observe a clear tendency that the preference to local neighborhoods increases with increasing node degree (which is known to correlate with poor NCP clusters and deep $k$ cores~\cite{leskovec2009community,jeub2015think,ASM13}).
The \textit{LINE} algorithm considers only one hop neighbors, and the assumption that only direct neighbors are relevant is very strong, especially for low degree nodes.

\subsubsection{Locality for more versus less peripheral classes}
Large graphs with flat NCP, especially with large and highly connected regions (with large deep \textit{k-cores}) are notably affected by random walk problems. 
For graphs with flat NCPs, the connectivity among nodes' \textit{relevant} neighbors is not much stronger than to the rest of the graph. 
Furthermore, the larger and deeper are graphs \textit{k-cores}, the more time random walks will spend in them. 
This affects the neighborhoods obtained by random walks for most nodes, since most parts of even large graphs can be reached within few steps.
Therefore, even if dense parts of the graph have high probabilities to be visited by global random walks, if the probabilities of single nodes in these components are low, then nodes from these components are not considered by \textsc{Lasagne} as neighbors.
Consequently, for the nodes from large deep \textit{k-cores}, the neighborhood will be restricted to the most \textit{relevant} core neighbors. 
Therefore, our method adapts to the structure of local neighborhood. 

To confirm this intuition, we used the Flickr network, a graph with flat NCP. Figure~\ref{fig:flickr_kcore} shows the fraction of nodes in different \textit{k-cores} of this graph. As can be seen in Figure\ref{fig:flickr_kcore}, the graph has large deep \textit{k-cores}, e.g., about 30\% of nodes are in the subgraph where each node has degree 100 or more. We expect random walk based methods to perform poorly on such a graph, especially if the similarity to neighbors outside of large deep \textit{k-cores} is important for the downstream task.

\begin{figure}[htb]
	\centering
  \includegraphics[width=0.65\textwidth]{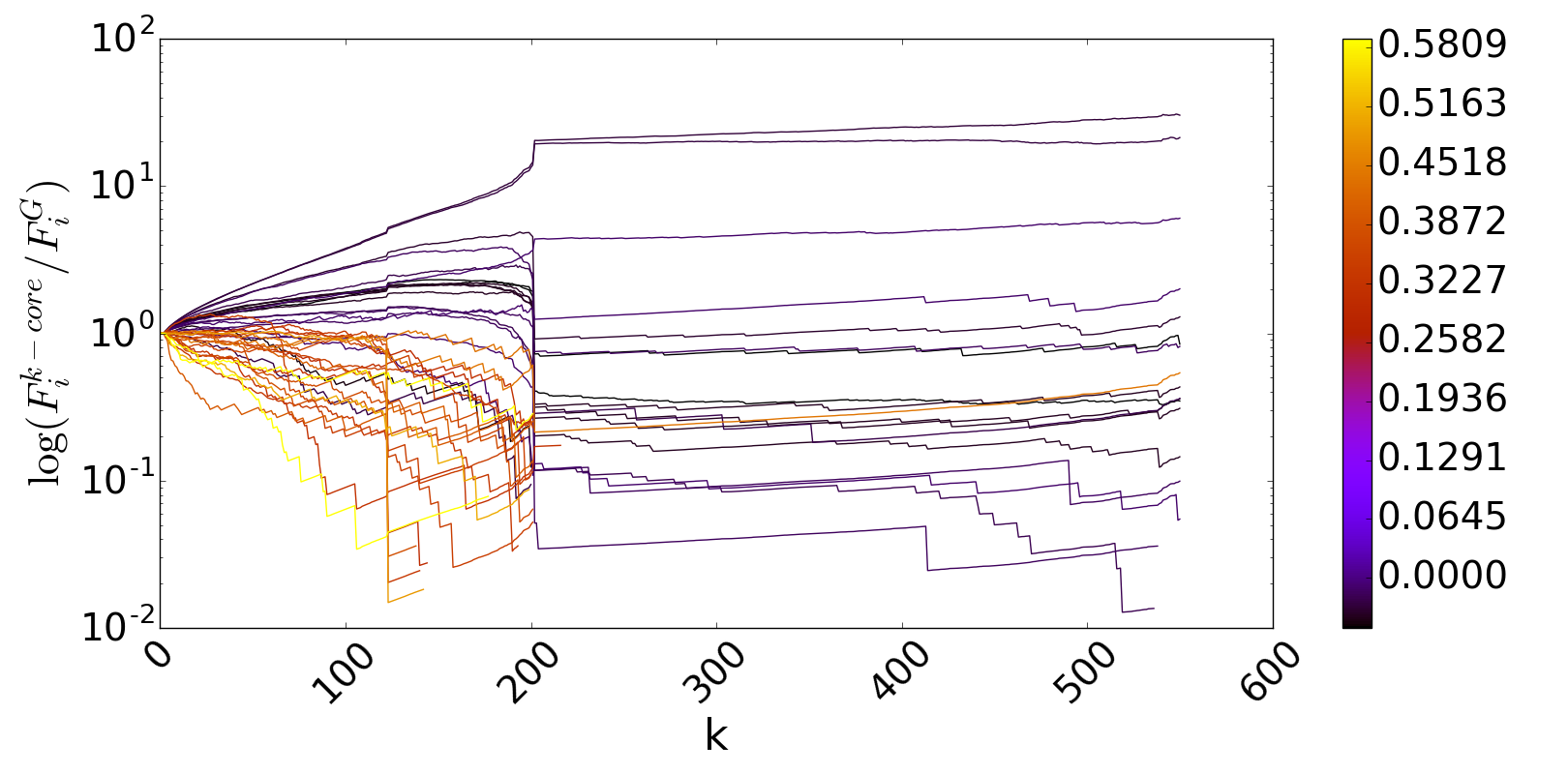}
	\caption{Each line depicts the class label distribution in k-cores with performance information for one class in the Flickr data. X-axis: k-core; Y-axis: log scaled proportion between fraction of class label $i$ within the k-core, i.e., $F_i^{k-core}$, and the fraction of this class label within the entire graph $G$, i.e., $F_i^{G}$; color code: absolute difference in $F_1$ score between \textsc{Lasagne} and the best random walk based method. For ease of presentation, the plot shows only the 20 classes where \textsc{Lasagne} reached the highest improvement as well as the 20 classes where the improvement was smallest.
  }
	\label{fig:class_results_k_cores}
\end{figure}

As multi-label classification is a common downstream task, Figure~\ref{fig:class_results_k_cores} provides empirical evidence that \textsc{Lasagne}'s embeddings overcome performance issues of previous embeddings.
In Figure~\ref{fig:class_results_k_cores}, each line stands for a class, and the color depicts the classification improvement of \textsc{Lasagne} over best previous method. 
Additionally, the plotted line shows the fraction of nodes with the corresponding class label in each \textit{k-core}, relative to the fraction of nodes with that label in the entire graph. 
When the fraction of class labels is zero, the line breaks. 
It can be clearly seen from the plot that \textsc{Lasagne} achieves the best improvement for classes with members outside of large \textit{k-cores} with high \textit{k}, i.e., for classes that are more peripheral.

\begin{figure}[t]
\begin{center}
    \subfigure[PPI]{
        \includegraphics[width =
        0.40\columnwidth]{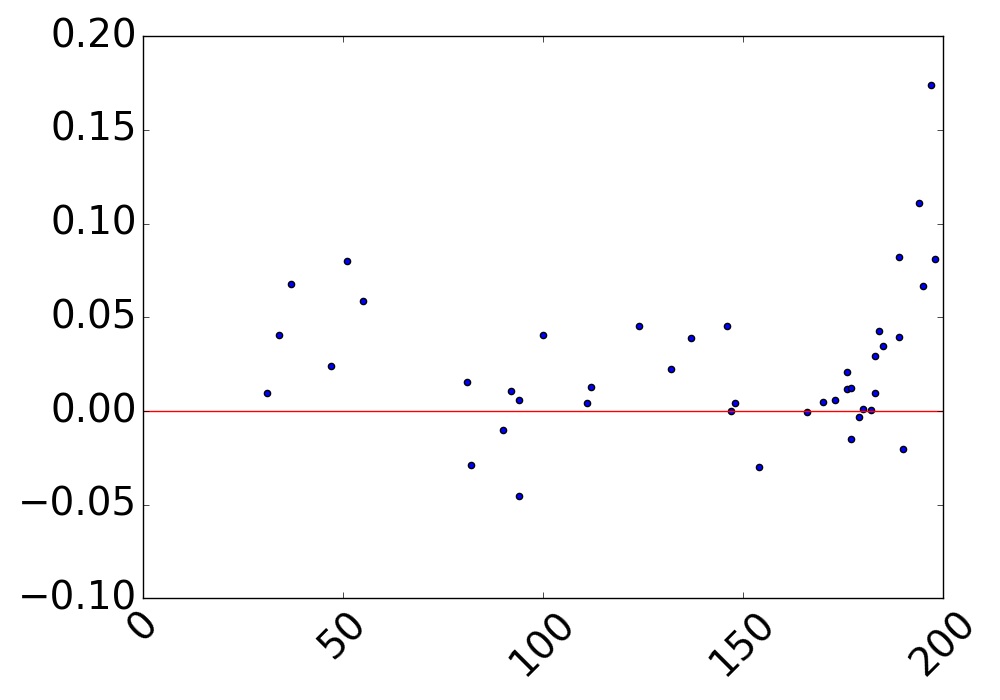}
  \label{fig:ppi_single_classes}
   } \subfigure[BlogCatalog]{
       \includegraphics[width =
       0.40\columnwidth]{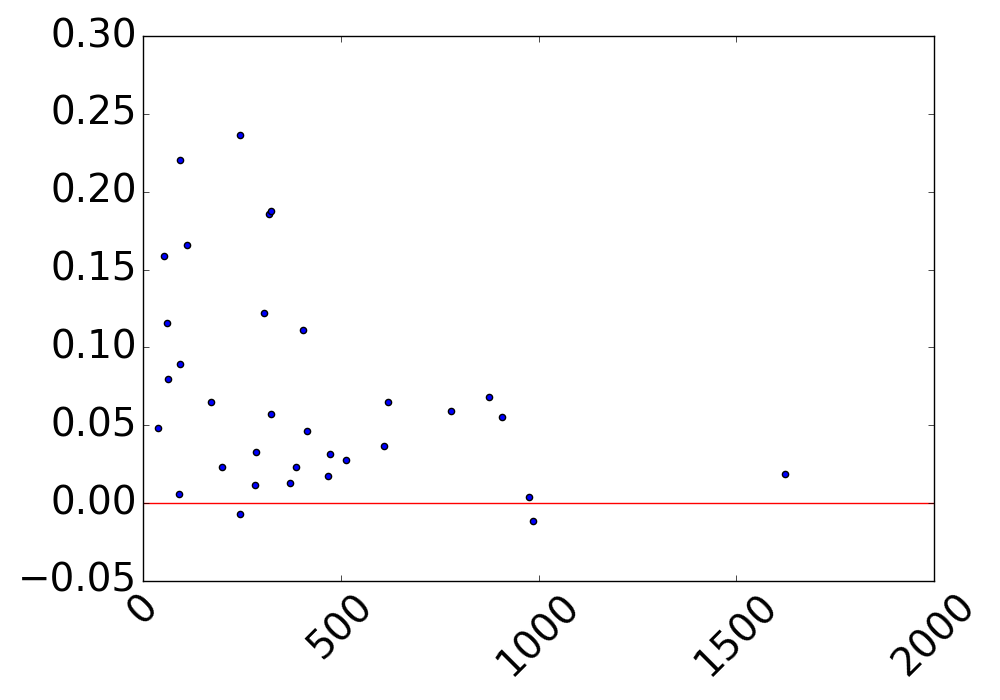}
       \label{fig:blogcatalog_single_classes}
   }
 \subfigure[IMDb Germany]{
       \includegraphics[width =
       0.40\columnwidth]{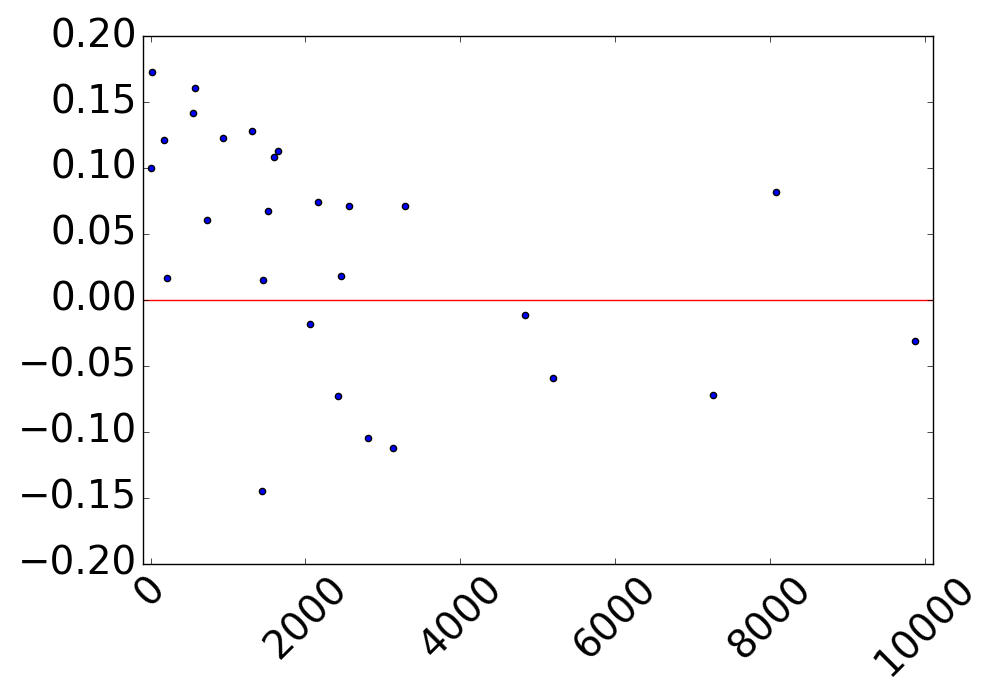}
  \label{fig:imdb_single_classes}
   } \subfigure[Flickr]{
       \includegraphics[width =
       0.40\columnwidth]{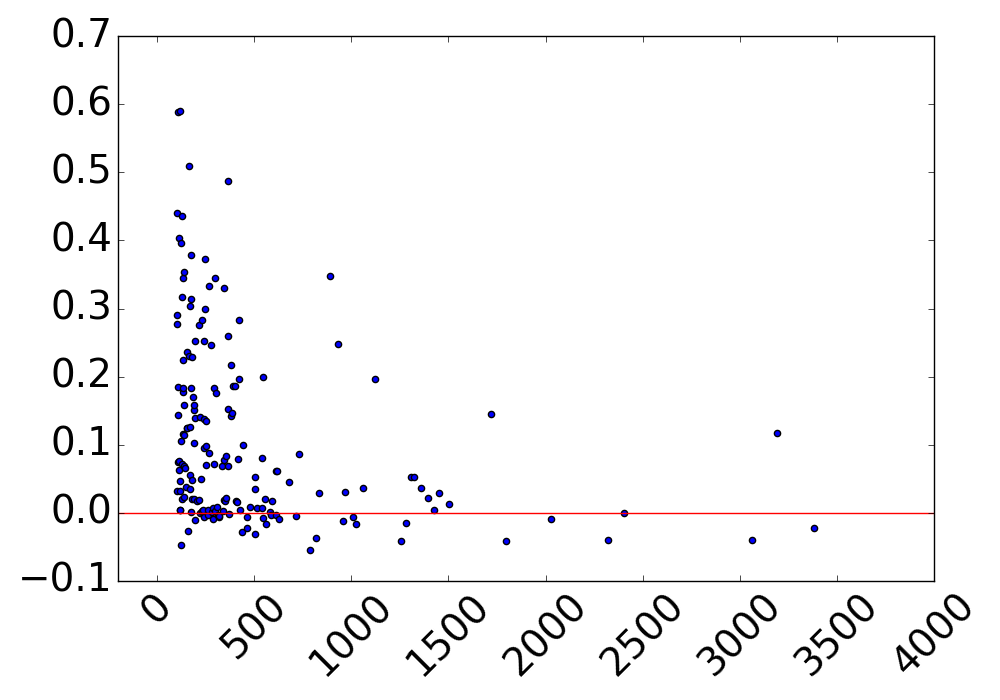}
       \label{fig:flickr_single_classes}
   } %
    \caption{Absolute differences in $F_1$ scores between \textsc{Lasagne} and the best random walk based method for single classes. One dot stands for one class. X-axes show the class size; Y-axes show the absolute difference in $F_1$ scores.}%
\label{fig:all_single_classes}
\end{center}
\end{figure}

Relatedly, we also expect that our method has better performance on the tasks which require very accurate determination of local neighborhoods. 
Small classes (in particular) need this, since nodes of such classes are very sensitive to irrelevant neighbors. 
This is due to the small number of nodes that belong to the same class. 
Figure~\ref{fig:all_single_classes} shows plots which visualize the improvement of \textsc{Lasagne} over random walk approaches for single classes. Please note different scale of X-axes for different datasets.
The improvements tend to be especially notable for the small classes, which confirms our claim. Due to its size all classes in PPI dataset are small.

\subsubsection{Distribution of training examples per node}
Another shortcoming of existing random walk based methods is the distribution of training examples per node that they generate. 
Since high-degree nodes are visited more often by random walks, there are more training examples for them.
Since small-degree nodes are visited much less often, they are underrepresented during training.
Due to the way in which locality is engineered into \textsc{Lasagne}, it solves this problem.

\begin{figure}[t]
    \subfigure[\textit{node2vec}]{
        \includegraphics[width =
        0.46\columnwidth]{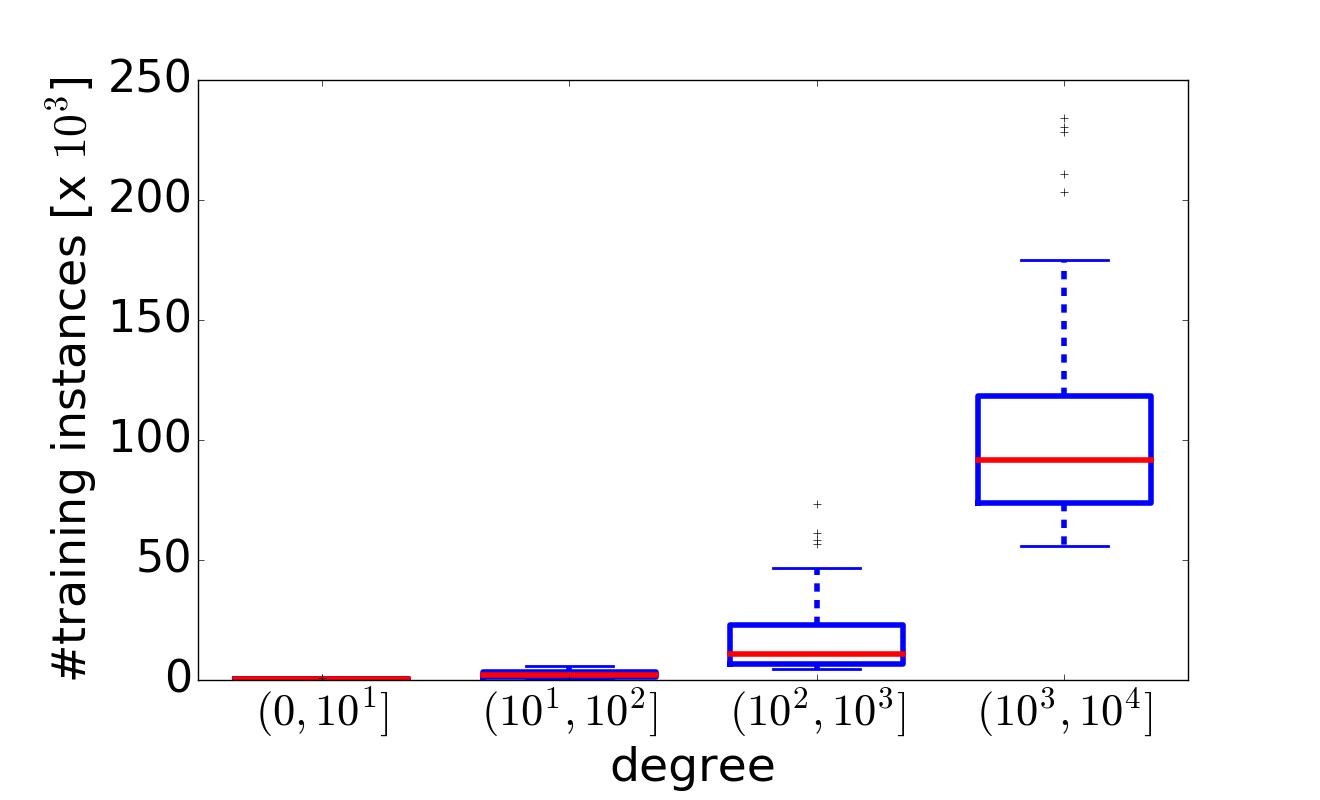}
  \label{fig:node2vec_training_instances}
   } \subfigure[\textsc{Lasagne}]{
       \includegraphics[width =
       0.46\columnwidth]{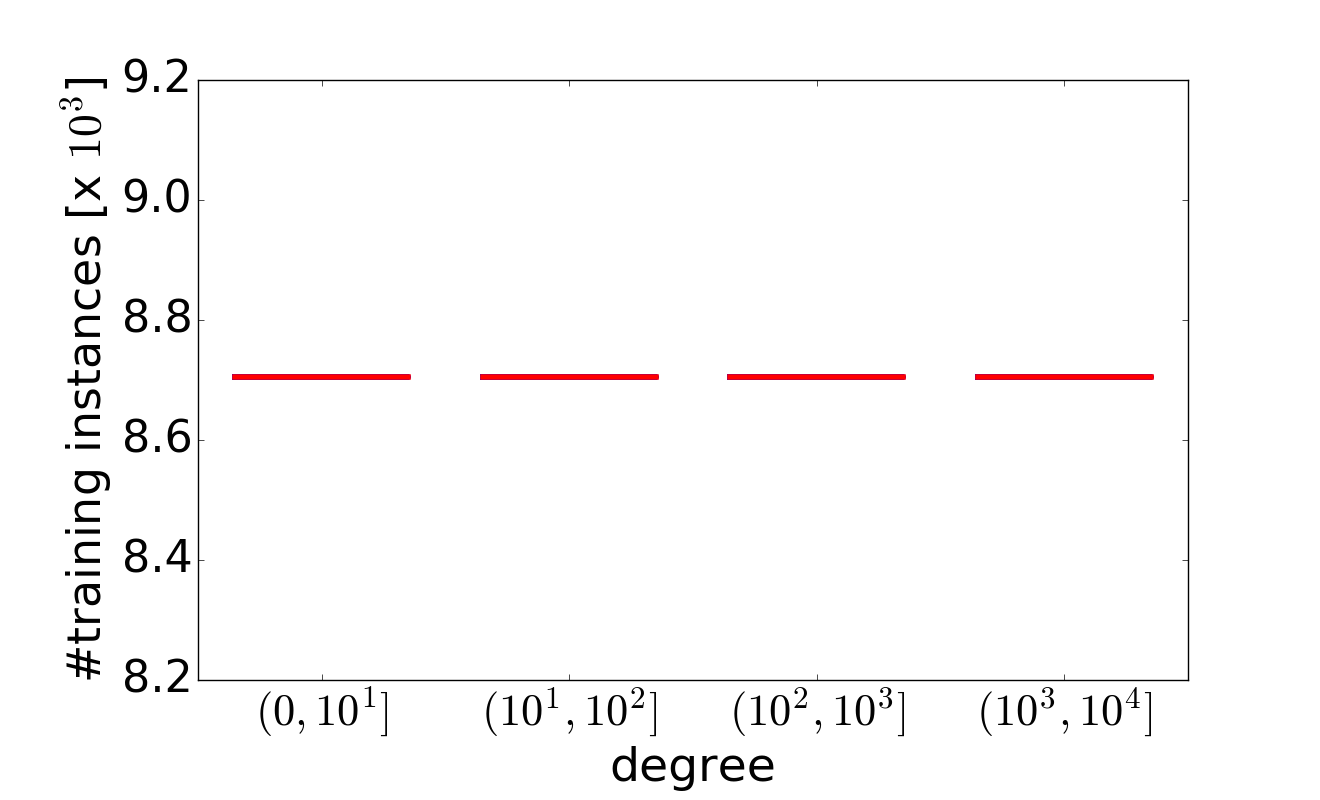}
       \label{fig:lasagne_training_instances}
   }
    \caption{The number of considered training instances for nodes of different degrees for \textit{node2vec} (left) and \textsc{Lasagne} (right). \textsc{Lasagne} allows us to control the number of training instances per node; we used 8700 for this plot.}%
    \label{fig:training_instances}
\end{figure}

To confirm this intuition, we run random walk based algorithms for different datasets and counted the number of training examples for a sample of nodes with different degrees. For each degree range 100 nodes were randomly sampled.
For all datasets we observed very similar distributions, also with different \textit{node2vec} parameters. An example is shown in Figure~\ref{fig:node2vec_training_instances}. 
As can be seen, the number of training examples for previous methods still strongly depends on node degrees. 
In contrast, when using \textsc{Lasagne}, the number of training examples per node can easily be controlled, in this case to be uniform, which prevents us from generating extremely unbalanced training sets, c.f. Figure~\ref{fig:lasagne_training_instances}.

\vspace{-1ex}  
\section{Conclusion}
\vspace{-1ex}  
\label{sec:conclusion}

We have proposed  \textsc{Lasagne}, an unsupervised learning algorithm to compute embeddings for the nodes of a graph.
The basic idea of \textsc{Lasagne} is to use an \textit{Approximate Personalized PageRank} algorithm to bias random walks more strongly to the local neighborhood of each node; and, thus, the embedding for a given node is more finely tuned to the local graph structure around that node than the embeddings from previous similar methods.
Our method performs particularly well for larger graphs that are not well-structured, e.g., that have flat NCPs and/or have many nodes in deep $k$-cores.
Our empirical evaluation has shown that our embeddings achieve superior prediction accuracy over competitors when used for multi-label classification in several different real-world networks.
Our empirical results also provide evidence justifying the reason for this improvement.
While \textsc{Lasagne} is primarily an exploratory tool, if one wants to use it in a more automated manner, then an important question will be how to automate the averaging of the APPR vectors over different values of the locality parameter.

\section*{Acknowledgements}
\noindent
This work was partially supported by Siemens, the Army Research Office, the Defense Advanced Research Projects Agency, and it was developed in cooperation with the Berkeley Institute of Data Science.
\bibliographystyle{IEEEtran}
\bibliography{bibliography}  %

\begin{thebibliography}{10}
\providecommand{\url}[1]{#1}
\csname url@samestyle\endcsname
\providecommand{\newblock}{\relax}
\providecommand{\bibinfo}[2]{#2}
\providecommand{\BIBentrySTDinterwordspacing}{\spaceskip=0pt\relax}
\providecommand{\BIBentryALTinterwordstretchfactor}{4}
\providecommand{\BIBentryALTinterwordspacing}{\spaceskip=\fontdimen2\font plus
\BIBentryALTinterwordstretchfactor\fontdimen3\font minus
  \fontdimen4\font\relax}
\providecommand{\BIBforeignlanguage}[2]{{%
\expandafter\ifx\csname l@#1\endcsname\relax
\typeout{** WARNING: IEEEtran.bst: No hyphenation pattern has been}%
\typeout{** loaded for the language `#1'. Using the pattern for}%
\typeout{** the default language instead.}%
\else
\language=\csname l@#1\endcsname
\fi
#2}}
\providecommand{\BIBdecl}{\relax}
\BIBdecl

\bibitem{maaten2008visualizing}
L.~v.~d. Maaten and G.~Hinton, ``Visualizing data using t-sne,'' \emph{Journal
  of Machine Learning Research}, vol.~9, no. Nov, pp. 2579--2605, 2008.

\bibitem{liben2007link}
D.~Liben-Nowell and J.~Kleinberg, ``The link-prediction problem for social
  networks,'' \emph{journal of the Association for Information Science and
  Technology}, vol.~58, no.~7, pp. 1019--1031, 2007.

\bibitem{aggarwal2011introduction}
C.~C. Aggarwal, ``An introduction to social network data analytics,'' in
  \emph{Social network data analytics}.\hskip 1em plus 0.5em minus 0.4em\relax
  Springer, 2011, pp. 1--15.

\bibitem{bhagat2011node}
S.~Bhagat, G.~Cormode, and S.~Muthukrishnan, ``Node classification in social
  networks,'' in \emph{Social network data analytics}.\hskip 1em plus 0.5em
  minus 0.4em\relax Springer, 2011, pp. 115--148.

\bibitem{perozzi2014deepwalk}
B.~Perozzi, R.~Al-Rfou, and S.~Skiena, ``Deepwalk: Online learning of social
  representations,'' in \emph{Proc. of the 20th ACM SIGKDD}, 2014, pp.
  701--710.

\bibitem{tang2015line}
J.~Tang, M.~Qu, M.~Wang, M.~Zhang, J.~Yan, and Q.~Mei, ``Line: Large-scale
  information network embedding,'' in \emph{Proc. of the 24th WWW}.\hskip 1em
  plus 0.5em minus 0.4em\relax ACM, 2015, pp. 1067--1077.

\bibitem{grover2016node2vec}
A.~Grover and J.~Leskovec, ``node2vec: Scalable feature learning for
  networks,'' in \emph{Proc. of the 22nd ACM SIGKDD}, 2016, pp. 855--864.

\bibitem{LLDM08_communities_CONF}
J.~Leskovec, K.~Lang, A.~Dasgupta, and M.~W. Mahoney, ``Statistical properties
  of community structure in large social and information networks,'' in
  \emph{WWW '08: Proceedings of the 17th International Conference on World Wide
  Web}, 2008, pp. 695--704.

\bibitem{leskovec2009community}
J.~Leskovec, K.~J. Lang, A.~Dasgupta, and M.~W. Mahoney, ``Community structure
  in large networks: Natural cluster sizes and the absence of large
  well-defined clusters,'' \emph{Internet Mathematics}, vol.~6, no.~1, pp.
  29--123, 2009.

\bibitem{LLM10_communities_CONF}
J.~Leskovec, K.~Lang, and M.~W. Mahoney, ``Empirical comparison of algorithms
  for network community detection,'' in \emph{WWW '10: Proceedings of the 19th
  International Conference on World Wide Web}, 2010, pp. 631--640.

\bibitem{jeub2015think}
L.~G. Jeub, P.~Balachandran, M.~A. Porter, P.~J. Mucha, and M.~W. Mahoney,
  ``Think locally, act locally: Detection of small, medium-sized, and large
  communities in large networks,'' \emph{Physical Review E}, vol.~91, no.~1, p.
  012821, 2015.

\bibitem{ASM13}
A.~B. Adcock, B.~D. Sullivan, and M.~W. Mahoney, ``Tree-like structure in large
  social and information networks,'' in \emph{Proc. of the 2013 IEEE ICDM},
  2013, pp. 1--10.

\bibitem{ASM16_IM}
------, ``Tree decompositions and social graphs,'' \emph{Internet Mathematics},
  vol.~12, no.~5, pp. 315--361, 2016.

\bibitem{andersen2006local}
R.~Andersen, F.~Chung, and K.~Lang, ``Local graph partitioning using pagerank
  vectors,'' in \emph{Proc. of the 47th IEEE FOCS}.\hskip 1em plus 0.5em minus
  0.4em\relax IEEE, 2006, pp. 475--486.

\bibitem{Shun:2016:PLG:2994509.2994522}
\BIBentryALTinterwordspacing
J.~Shun, F.~Roosta-Khorasani, K.~Fountoulakis, and M.~W. Mahoney, ``Parallel
  local graph clustering,'' \emph{Proc. VLDB Endow.}, vol.~9, no.~12, pp.
  1041--1052, Aug. 2016. [Online]. Available:
  \url{http://dx.doi.org/10.14778/2994509.2994522}
\BIBentrySTDinterwordspacing

\bibitem{tenenbaum2000global}
J.~B. Tenenbaum, V.~De~Silva, and J.~C. Langford, ``A global geometric
  framework for nonlinear dimensionality reduction,'' \emph{science}, vol. 290,
  no. 5500, pp. 2319--2323, 2000.

\bibitem{roweis2000nonlinear}
S.~T. Roweis and L.~K. Saul, ``Nonlinear dimensionality reduction by locally
  linear embedding,'' \emph{Science}, vol. 290, no. 5500, pp. 2323--2326, 2000.

\bibitem{belkin2001laplacian}
M.~Belkin and P.~Niyogi, ``Laplacian eigenmaps and spectral techniques for
  embedding and clustering.'' in \emph{NIPS}, vol.~14, 2001, pp. 585--591.

\bibitem{cao2015grarep}
S.~Cao, W.~Lu, and Q.~Xu, ``Grarep: Learning graph representations with global
  structural information,'' in \emph{Proceedings of the 24th ACM International
  on Conference on Information and Knowledge Management}.\hskip 1em plus 0.5em
  minus 0.4em\relax ACM, 2015, pp. 891--900.

\bibitem{yang2016revisiting}
Z.~Yang, W.~Cohen, and R.~Salakhudinov, ``Revisiting semi-supervised learning
  with graph embeddings,'' in \emph{Proc. of The 33rd ICDM}, 2016, pp. 40--48.

\bibitem{scarselli2009graph}
F.~Scarselli, M.~Gori, A.~C. Tsoi, M.~Hagenbuchner, and G.~Monfardini, ``The
  graph neural network model,'' \emph{IEEE Transactions on Neural Networks},
  vol.~20, no.~1, pp. 61--80, 2009.

\bibitem{yan2005graph}
S.~Yan, D.~Xu, B.~Zhang, and H.-J. Zhang, ``Graph embedding: A general
  framework for dimensionality reduction,'' in \emph{Proc. of IEEE CVPR},
  vol.~2.\hskip 1em plus 0.5em minus 0.4em\relax IEEE, 2005, pp. 830--837.

\bibitem{li2014lrbm}
K.~Li, J.~Gao, S.~Guo, N.~Du, X.~Li, and A.~Zhang, ``Lrbm: A restricted
  boltzmann machine based approach for representation learning on linked
  data,'' in \emph{Proc. of IEEE ICDM}.\hskip 1em plus 0.5em minus 0.4em\relax
  IEEE, 2014, pp. 300--309.

\bibitem{li2016gated}
Y.~Li, D.~Tarlow, M.~Brockschmidt, and R.~Zemel, ``Gated graph sequence neural
  networks,'' in \emph{ICLR}, 2016.

\bibitem{tian2014learning}
F.~Tian, B.~Gao, Q.~Cui, E.~Chen, and T.-Y. Liu, ``Learning deep
  representations for graph clustering.'' in \emph{AAAI}, 2014, pp. 1293--1299.

\bibitem{kipf2016semi}
T.~N. Kipf and M.~Welling, ``Semi-supervised classification with graph
  convolutional networks,'' \emph{arXiv preprint arXiv:1609.02907}, 2016.

\bibitem{mikolov2013efficient}
T.~Mikolov, K.~Chen, G.~Corrado, and J.~Dean, ``Efficient estimation of word
  representations in vector space,'' \emph{arXiv preprint arXiv:1301.3781},
  2013.

\bibitem{le2014distributed}
Q.~V. Le and T.~Mikolov, ``Distributed representations of sentences and
  documents.'' in \emph{ICML}, vol.~14, 2014, pp. 1188--1196.

\bibitem{gittens2017skip}
A.~Gittens, D.~Achlioptas, and M.~W. Mahoney, ``Skip-gram-zipf+ uniform= vector
  additivity,'' in \emph{Proc. of ACL}, vol.~1, 2017, pp. 69--76.

\bibitem{levy2014dependency}
O.~Levy and Y.~Goldberg, ``Dependency-based word embeddings.'' in \emph{ACL
  (2)}, 2014, pp. 302--308.

\bibitem{page1999pagerank}
L.~Page, S.~Brin, R.~Motwani, and T.~Winograd, ``The pagerank citation ranking:
  bringing order to the web.'' 1999.

\bibitem{jeh2003scaling}
G.~Jeh and J.~Widom, ``Scaling personalized web search,'' in \emph{Proc. of the
  12th WWW}.\hskip 1em plus 0.5em minus 0.4em\relax ACM, 2003, pp. 271--279.

\bibitem{berkhin2006bookmark}
P.~Berkhin, ``Bookmark-coloring algorithm for personalized pagerank
  computing,'' \emph{Internet Mathematics}, vol.~3, no.~1, pp. 41--62, 2006.

\bibitem{knuth1969art}
D.~Knuth, ``The art of computer programming: Vol 2/seminumerical algorithms,
  chapter 3: Random numbers,'' 1969.

\bibitem{ji2016parallelizing}
S.~Ji, N.~Satish, S.~Li, and P.~Dubey, ``Parallelizing word2vec in shared and
  distributed memory,'' \emph{arXiv preprint arXiv:1604.04661}, 2016.

\bibitem{breitkreutz2008biogrid}
B.-J. Breitkreutz, C.~Stark, T.~Reguly, L.~Boucher, A.~Breitkreutz,
  M.~Livstone, R.~Oughtred, D.~H. Lackner, J.~B{\"a}hler, V.~Wood
  \emph{et~al.}, ``The biogrid interaction database: 2008 update,''
  \emph{Nucleic acids research}, vol.~36, no. suppl 1, pp. D637--D640, 2008.

\bibitem{tang2009relational}
L.~Tang and H.~Liu, ``Relational learning via latent social dimensions,'' in
  \emph{Proc. of the 15th ACM SIGKDD}.\hskip 1em plus 0.5em minus 0.4em\relax
  ACM, 2009, pp. 817--826.

\bibitem{imdb2016the}
IMDb, ``The internet movie database,'' in \emph{http://www.imdb.com/}, 2016,
  last accessed: 2016-11-22.

\bibitem{shun2013ligra}
J.~Shun and G.~E. Blelloch, ``Ligra: a lightweight graph processing framework
  for shared memory,'' in \emph{ACM SIGPLAN Notices}, vol.~48, no.~8.\hskip 1em
  plus 0.5em minus 0.4em\relax ACM, 2013, pp. 135--146.

\bibitem{abadi2016tensorflow}
M.~Abadi, P.~Barham, J.~Chen, Z.~Chen, A.~Davis, J.~Dean, M.~Devin,
  S.~Ghemawat, G.~Irving, M.~Isard \emph{et~al.}, ``Tensorflow: A system for
  large-scale machine learning,'' in \emph{Proc. of the 12th USENIX OSDI.
  Savannah, Georgia, USA}, 2016.

\bibitem{snapnets}
J.~Leskovec and A.~Krevl, ``{SNAP Datasets}: {Stanford} large network dataset
  collection,'' \url{http://snap.stanford.edu/data}, Jun. 2014.

\end{thebibliography}

\end{document}